%% file: main_file.tex
\documentclass[longauth]{aa}  

\usepackage{graphicx}
\usepackage{txfonts}
\usepackage{comment}
\usepackage[dvipsnames]{xcolor}
\pdfoutput=1
\definecolor{mypink}{RGB}{219,48,122}

\begin{document} 

\title{Strong detection of the CMB lensing $\times$ galaxy weak lensing cross-correlation from ACT-DR4, \textit{Planck} Legacy and KiDS-1000}

\author{Naomi Clare Robertson\inst{1,2,3}\fnmsep\thanks{ncr@ast.cam.ac.uk}, David Alonso\inst{3}, Joachim Harnois-D\'eraps\inst{4,5}, Omar Darwish\inst{6}, Arun Kannawadi\inst{7},
Alexandra Amon\inst{8}, 
Marika Asgari\inst{5}, 
Maciej Bilicki\inst{9},
Erminia Calabrese\inst{10},
Steve~K.~Choi\inst{11,12},
Mark J. Devlin\inst{13},
Jo Dunkley\inst{7,14},
Andrej Dvornik\inst{15},
Thomas Erben\inst{16}, 
Simone Ferraro\inst{17,18},
Maria Cristina Fortuna\inst{19}
Benjamin Giblin\inst{5},
Dongwon Han\inst{20},
Catherine Heymans\inst{5,16},
Hendrik Hildebrandt\inst{15}
J. Colin Hill\inst{21,22},
Matt Hilton\inst{23,24},
Shuay-Pwu P. Ho\inst{25},
Henk Hoekstra\inst{19},
Johannes Hubmayr\inst{26},
Jack Hughes\inst{27},
Benjamin Joachimi\inst{28},
Shahab Joudaki\inst{3,29},
Kenda Knowles\inst{23},
Konrad Kuijken\inst{19},
Mathew S. Madhavacheril\inst{30},
Kavilan Moodley\inst{23,24},
Lance Miller\inst{3},
Toshiya Namikawa\inst{6},
Federico Nati\inst{31},
Michael D. Niemack\inst{11,12},
Lyman A. Page\inst{14},
Bruce Partridge\inst{32},
Emmanuel Schaan\inst{17,18},
Alessandro Schillaci\inst{33},
Peter Schneider\inst{16},
Neelima Sehgal\inst{20},
Blake D. Sherwin\inst{6,2},
Crist\'obal \ Sif\'on\inst{34},
Suzanne T. Staggs\inst{14},
Tilman Tröster\inst{5},
Alexander van Engelen\inst{35},
Edwin Valentijn\inst{36},
Edward J. Wollack\inst{37},
Angus H. Wright\inst{15},
Zhilei Xu\inst{13,38}}

\institute{
Institute of Astronomy, University of Cambridge, Madingley Road, Cambridge, CB3 0HA
\and
Kavli Institute for Cosmology Cambridge, Madingley Road, Cambridge CB3 0HA
\and
Sub-department of Astrophysics, University of Oxford, Keble Road, Oxford OX1 3RH, UK 
\and
Astrophysics Research Institute, Liverpool John Moores University, 146 Brownlow Hill, Liverpool L3 5RF
\and
Institute for Astronomy, University of Edinburgh, Royal Observatory, Blackford Hill, Edinburgh, EH9 3HJ, UK 
\and
Department of Applied Mathematics and Theoretical Physics, University of Cambridge, Wilberforce Road, Cambridge CB3 0WA
\and
Department of Astrophysical Sciences, Princeton University, 4 Ivy Lane, Princeton, NJ 08544, USA
\and
Kavli Institute for Particle Astrophysics \& Cosmology, P.O. Box 2450, Stanford University, Stanford, CA 94305, USA 
\and
Center for Theoretical Physics, Polish Academy of Sciences, al. Lotników 32/46, 02-668, Warsaw, Poland
\and
School of Physics and Astronomy, Cardiff University, The Parade, Cardiff, CF24 3AA, UK
\and
Department of Physics, Cornell University, Ithaca, NY, USA 14853
\and
Department of Astronomy, Cornell University, Ithaca, NY 14853, USA
\and
Department of Physics and Astronomy, University of Pennsylvania, 209 South 33rd Street, Philadelphia, PA 19104, USA
\and
Joseph Henry Laboratories of Physics, Jadwin Hall, Princeton University, Princeton, NJ 08544, USA 
\and
Ruhr University Bochum, Faculty of Physics and Astronomy, Astronomical Institute (AIRUB), German Centre for Cosmological Lensing, 44780 Bochum, Germany
\and
Argelander-Institut für Astronomie, Auf dem Hügel 71, 53121 Bonn, Germany
\and
Lawrence Berkeley National Laboratory, One Cyclotron Road, Berkeley, CA 94720, USA
\and
Berkeley Center for Cosmological Physics, UC Berkeley, CA 94720, USA
\and
Leiden Observatory, Leiden University, P.O.Box 9513, 2300RA Leiden, The Netherlands
\and
Physics and Astronomy Department, Stony Brook University, Stony Brook, NY 11794
\and
Department of Physics, Columbia University, New York, NY, USA 10027
\and
Center for Computational Astrophysics, Flatiron Institute, New York, NY, USA 10010
\and
Astrophysics Research Centre, University of KwaZulu-Natal, Westville Campus, Durban 4041, South Africa
\and
School of Mathematics, Statistics \& Computer Science, University of KwaZulu-Natal, Westville Campus, Durban4041, South Africa
\and
Department of Physics, Stanford University, Stanford, CA, USA 94305-4085
\and
NIST Quantum Sensors Group, 325 Broadway, Boulder, CO 80305
\and
Department of Physics and Astronomy, Rutgers, the State University of New Jersey, 136 Frelinghuysen Road, Piscataway, NJ 08854-8019, USA
\and
Department of Physics and Astronomy, University College London, Gower Street, London WC1E 6BT, UK
\and
Waterloo Centre for Astrophysics, University of Waterloo, 200 University Ave W, Waterloo, ON N2L 3G1, Canada
\and
Centre for the Universe, Perimeter Institute for Theoretical Physics, Waterloo, ON, N2L 2Y5, Canada
\and
Department of Physics, University of Milano-Bicocca, Piazza della Scienza 3, 20126 Milano (MI), Italy
\and
Department of Physics and Astronomy, Haverford College,Haverford, PA, USA 19041
\and
Department of Physics, California Institute of Technology, Pasadena, CA 91125, USA
\and
Instituto de Física, Pontificia Universidad Católica de Valparaíso, Casilla 4059, Valpara íso, Chile
\and
School of Earth and Space Exploration, Arizona State Uni- versity, Tempe, AZ, USA 85287
\and
Kapteyn Institute, University of Groningen, PO Box 800, NL 9700 AV Groningen
\and
NASA/Goddard Space Flight Center, Greenbelt, MD 20771, USA
\and
MIT Kavli Institute, Massachusetts Institute of Technology, 77 Massachusetts Avenue, Cambridge, MA 02139, USA
}
\date{Received XXXX; accepted YYYY}

 
\abstract{We measure the cross-correlation between galaxy weak lensing data from the Kilo Degree Survey (KiDS-1000, DR4) and cosmic microwave background (CMB) lensing data from the Atacama Cosmology Telescope (ACT, DR4) and the \textit{Planck} Legacy survey. We use two samples of source galaxies, selected with photometric redshifts, $(0.1<z_{\rm B}<1.2)$ and $(1.2<z_{\rm B}<2)$, which produce a combined detection significance of the CMB lensing/weak galaxy lensing cross-spectrum of $7.7\sigma$. With the lower redshift galaxy sample, for which the cross-correlation is
detected at a significance of $5.3\sigma$, we present joint cosmological constraints on the matter density parameter, $\Omega_{\rm m}$, and the matter fluctuation amplitude parameter, $\sigma_8$, marginalising over three nuisance parameters that model our uncertainty in the redshift and shear calibration, and the intrinsic alignment of galaxies. We find our measurement to be consistent with the best-fitting flat $\Lambda$CDM cosmological models from both \textit{Planck} and KiDS-1000. We demonstrate the capacity of CMB-weak lensing cross-correlations to set constraints on either the redshift or shear calibration, by analysing a previously unused high-redshift KiDS galaxy sample $(1.2<z_{\rm B}<2)$, with the cross-correlation detected at a significance of $7\sigma$. This analysis provides an independent assessment for the accuracy of redshift measurements in a regime that is challenging to calibrate directly owing to known incompleteness in spectroscopic surveys.}

    \titlerunning{ACT/\textit{Planck} lensing$\times$KiDS-1000 lensing}
    \authorrunning{Robertson \& the KiDS and ACT Collaborations et al.}
    \maketitle

%
\section{Introduction}
\label{introduction}
Recent measurements from cosmic microwave background (CMB) experiments, galaxy redshift surveys, weak gravitational lensing, Type-Ia Supernovae, Cepheid Variable stars and time delays from strong lensing systems have put stringent constraints on the parameters that describe the standard model of cosmology \citep[e.g.,][]{SPT2018,Freedman2019,Riess2019,Abbott2018,Hikage2019,Planck2018,  Wong2020,Heymans2020,eBOSS2020,Aiola2019,Choi2020}. While the flat $\Lambda$CDM model is favoured in most of these surveys when considered individually, some tensions are observed in their combination -- notably tensions with the Hubble parameter, $H_0$  \citep{Riess2019, Wong2020} and $\sigma_8$, the amplitude of matter fluctuations \citep{Abbott2018,Hikage2019,Heymans2020}. Determining whether these arise from newly discovered phenomena or residual systematic effects is currently one of the key scientific goals in the field of cosmology, and a number of techniques have been proposed to approach this problem.

In recent years there have been many studies combining different tracers of the density field. These not only improve cosmological model constraints, but also help test various astrophysical and instrumental systematic effects which can limit our analyses and potentially cause tensions between parameters. For example, cross-correlating CMB lensing with populations of galaxies \citep[for example][]{Smith2007,Hirata2008,Bleem2012,Allison2015,Bianchini2015,Giannantonio2016,Peacock2018,Omori2018b,Krolewski2020, Darwish2019}, quasars \citep{Sherwin2012,Geach2013}, galaxy groups \citep{Madhavacheril2015} and clusters \citep{Baxter2018,Madhavacheril2020} can be used to measure the bias of these foreground tracers as well as the growth factor. Cross-correlating CMB lensing with the Cosmic Infrared Background \citep[CIB,][]{Holder2013,PlanckCIB,vanEngelen2015,CIBmap}, or correlating galaxy or CMB lensing with the thermal Sunyaev-Zel'dovich effect \citep{vanWaerbeke2014,Hill2014,Hojjati2017} informs us about the relationship between dark matter and baryons. 

In this study we measure the cross-correlation between galaxy weak lensing and CMB lensing, which both probe an overlapping volume of the underlying matter density field, and are therefore correlated. The galaxy weak lensing kernel peaks at around a redshift of 0.5 for current data sets, while the CMB lensing kernel is broad and peaks at a redshift of around 2 -- thus by measuring the cross-spectrum between these two data sets we can directly study the mass distribution at an intermediate redshift of around 1, purely from gravitational lensing \citep{bartelmann,lewis}.

In parallel with constraining cosmology, this particular cross-correlation measurement can help calibrate residual biases present in galaxy weak lensing \citep{Hu2002,Vallinotto2013}. These include potential errors in the shear calibration correction, $m$, (also described in the literature as the responsivity), that corrects for the impact of pixel-noise, object selection bias and galaxy blending \citep[see discussion in][and references therein]{Kannawadi2019}, and the source redshift calibration, that corrects for the imperfect match between photometric and spectroscopic redshifts.  Any error in the shear and/or redshift calibration will be degenerate with the normalisation of the matter power spectrum. As the CMB lensing kernel is defined by the underlying cosmology, cross-correlation analyses can potentially unearth unknown systematic errors, providing an independent validation of the more standard calibration methodology using mock catalogues and image simulations.

The first measurement of this kind combined the CMB lensing data from the Atacama Cosmology Telescope (ACT) with galaxy weak lensing data from the Canada-France-Hawaii Telescope Stripe 82 Survey \citep[CS82,][]{Hand2015}. There have been several subsequent measurements using the Canada-France Hawaii Telescope Lensing Survey, Red Cluster Survey Lensing, Kilo Degree Survey (KiDS), Dark Energy Survey, or Hyper Suprime-Cam galaxy weak lensing surveys combined with the \textit{Planck}, ACT, South Pole Telescope or POLARBEAR CMB lensing data \citep{Liu2015,Kirk2016,Harnois2016,Harnois2017,Miyatake2017,Omori2018,Namikawa2019,Marques2020}. Little improvement in the signal-to-noise ratio (SNR) has been observed between these measurements, mostly due either to the small overlapping area between the data sets or to the large residual noise in the CMB lensing maps; both considerations presently limit the achieved SNR. Ground-based CMB experiments that have lower noise and higher resolution than \textit{Planck} have recently increased their overlap with galaxy weak lensing surveys (e.g., SPT\footnote{https://pole.uchicago.edu} and ACT\footnote{https://act.princeton.edu}), so significant improvements are expected.

For the current generation of low SNR measurements, it is common to compress the cross-correlation measurement into an estimation of a single amplitude parameter $A$, which is equal to unity if the data are consistent with a $\Lambda$CDM model at a chosen cosmology. For the majority of these previous measurements, the amplitude $A$ was calibrated on the \textit{Planck} best-fit cosmology and found to be lower than expected \citep{Hand2015,Liu2015,Kirk2016,Harnois2016,Harnois2017,Marques2020}. There has been some discussion about whether this could be caused by the effect of intrinsic alignment of galaxies \citep{Hall2014,Chisari2015,Larsen2016}, which can imitate and conceal the true lensing signal \citep[see][and references therein]{Kirk2015}.
The cross-correlation of the intrinsic alignments of galaxies with the CMB lensing convergence produces a negative power spectrum -- analogous to the term describing the correlation between the intrinsic alignment and lensing signal in cosmic shear surveys -- resulting in a reduction in the cross-correlation amplitude \citep{Hirata2004}. 

\citet{Hall2014} found that for a sample with a redshift distribution similar to that of CS82, which has a median redshift of $\sim 0.6$, intrinsic alignment leads to a 15-20\% decrease in the amplitude of the cross-correlation signal. It should be noted, however, that this modification is dependent on the shape of the redshift distribution and the type of source galaxies sampled by the galaxy weak lensing survey. Furthermore, \citet{Chisari2015} showed this effect is dependent on the type of galaxy, with red galaxies leading to a shift of around 10\% in the amplitude and blue galaxies showing no alignment. \citet{Hall2014} made more extreme assumptions about the galaxy population and so should be considered an upper limit. While both of these investigations displayed that deeper surveys would have less contamination, a better understanding of the high-redshift intrinsic alignment will be required to validate these conclusions.

Beyond the effects of intrinsic alignments, \citet{Harnois2016} showed that baryonic feedback suppresses the cross-correlation power spectrum at large $\ell$-bins, and massive neutrinos are also known to affect the amplitude of the signal \citep[for example see][more broadly]{vanDaalen2011,Semboloni2011,Springel2018}. To date, these two effects are sub-dominant given the SNR of the measurements, but their proper modelling will be required for the upcoming analyses of Stage-IV galaxy lensing \citep{Harnois2014,Chisari2019} and CMB experiments \citep{Natarajan2014,Chung2019}. 

In this analysis, we measure the CMB lensing$\times$galaxy weak lensing power spectrum using the latest CMB lensing data from ACT and \textit{Planck}, and the fourth data release of the KiDS weak lensing observations, covering an area of more than 1000 deg$^2$ and achieving an unprecedented detection level of $7.7\sigma$. The data sets are detailed in Section \ref{sec:data}. Our method for measuring and modelling the cross-correlation signal is described in Section \ref{sec:method} and we present our results in Section \ref{sec:results}. We then perform an MCMC analysis to fit for $\Omega_{\rm m}$ and $\sigma_8$ while marginalising over the known galaxy weak lensing systematic biases. We also calibrate a higher redshift galaxy sample in the KiDS data. We conclude with a summary of our analysis in Section \ref{discussion}.

\section{Data}
\label{sec:data}
\subsection{Galaxy Weak Lensing}
KiDS is a recently completed optical wide-field imaging survey carried out with the OmegaCAM camera mounted on the VLT Survey Telescope. It primarily targets the weak gravitational lensing signal of galaxies with a mean redshift of around $z=0.7$. The final data release of KiDS will consist of 1350 deg$^2$ split across two patches of sky -- one along the equator (KiDS North, or KiDS-N) that benefits from overlapping with the GAMA and SDSS spectroscopic surveys as well as a single pointing in the COSMOS15 field \citep{cosmos15}, and the other in the south (KiDS South, or KiDS-S). KiDS observes the sky in four bands $u,g,r,i$, and the $r$-band data, having the best seeing conditions, is used for estimating the shapes of the galaxies. By design, the same area of the sky has also been observed by the VISTA Kilo-degree Infrared Galaxy survey (VIKING) using the VISTA telescope in the $Z,Y,J,H$ and $K_s$ bands \citep{Edge2013}, thus making KiDS a 9-band imaging survey \citep{Wright2019}. 

\begin{figure}
	\includegraphics[width=\columnwidth]{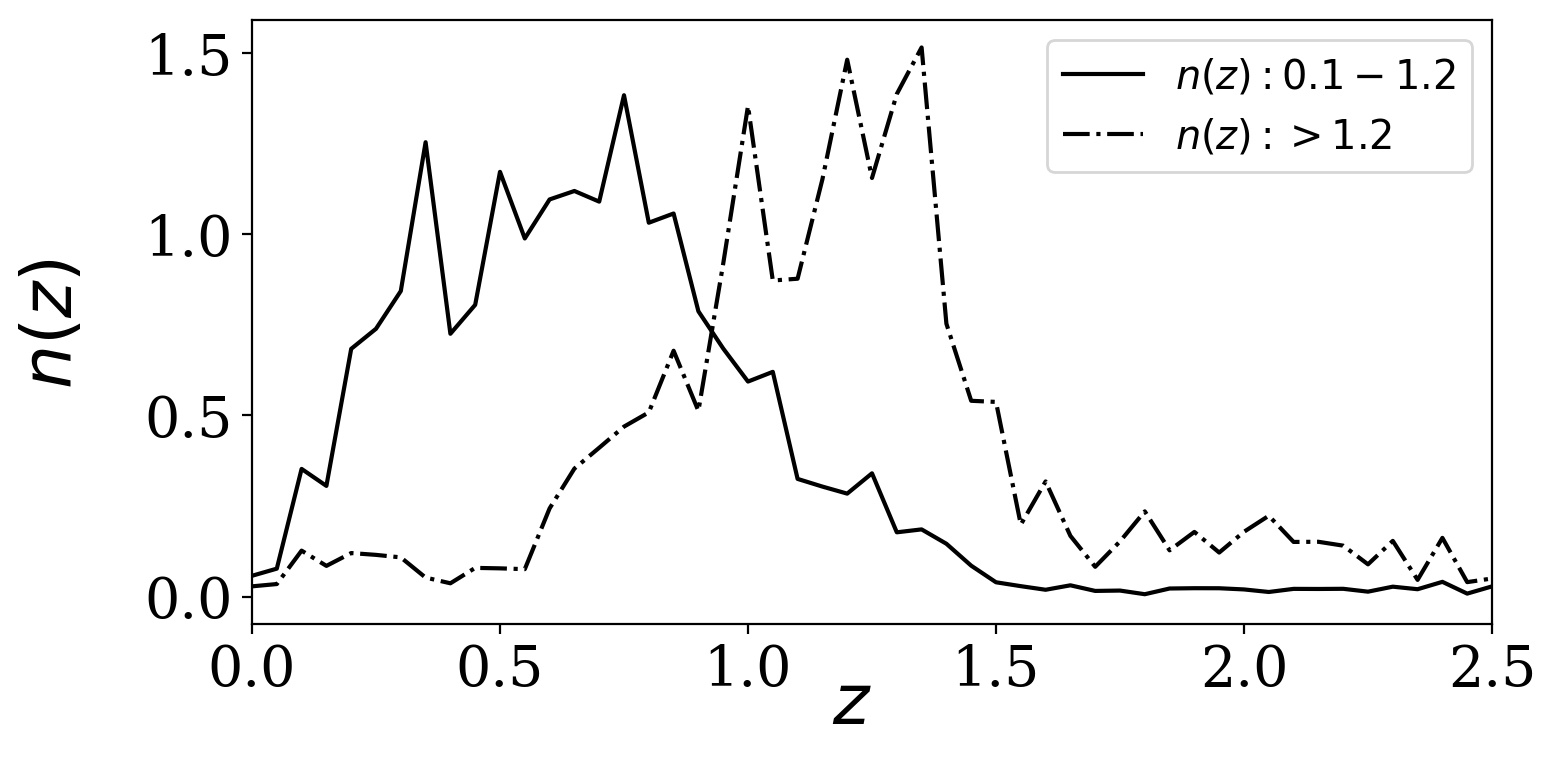}
    \caption{The redshift distributions, as estimated for the full KiDS-1000 galaxy sample, with a photometric redshift selection $(0.1<z_{\rm B}<1.2)$ (solid black) and $(1.2<z_{\rm B}<2)$ (dot-dashed black) \citep{Hildebrandt2020}. Both redshift distributions are derived using the DIR method \citep{Hildebrandt2018}.}
    \label{fig:nz}
\end{figure}

The measurements presented in this paper are based on the fourth KiDS data release (Kuijken et al 2019, hereafter KiDS-1000), which spans 1006 deg$^2$ with accurately calibrated source redshift distributions out to a photometric redshift $z_{\rm B} < 1.2$ \citep{Wright2020a}. This deep and wide-field data results in a significant overlap in terms of the galaxy weak lensing and CMB lensing window function, which improves the detection significance of our measurement in comparison to previous studies. The increased homogeneity in the sky coverage of KiDS-1000 also makes it ideal for cross-correlation measurements, as the footprint contains fewer sharp features. 

The shape measurements are carried out with the {\it lens}fit pipeline \citep{Miller2013}, calibrated using simulations presented in \citet{Kannawadi2019} and detailed in \citet{Giblin2020}, where residual shear-related systematics are shown to have a negligible impact on the cosmological inference with two-point statistics. Our cross-correlation measurement is even less sensitive to these, hence their conclusions also apply here.  

The redshift distribution used in this analysis was derived in the earlier KiDS+VIKING (KV450) analysis by \citet{Hildebrandt2018} using the weighted direct calibration (DIR) method. This method takes a sub-sample of the KiDS galaxies for which spectroscopic redshift information is available, augmented with a $k$NN re-weighting technique as detailed in \citet{Lima2008}. The redshift distribution, shown in Fig. \ref{fig:nz}, is produced by applying the DIR method to samples of galaxies binned by their $z_{\rm B}$ value -- the peak of the redshift posterior produced by the Bayesian photometric redshift BPZ code \citep{Benitez2000}, estimated using the 9-band photometry -- and weighted with the {\it lens}fit weights. This method produces a well-calibrated redshift distribution for galaxies with a BPZ-redshift in the range $(0.1<z_{\rm B}<1.2)$ that was used in the KV450 cosmic shear analyses, and here as well. 

In this analysis we choose a different redshift calibration approach than that adopted in the KiDS-1000 cosmology analyses \citep{Asgari2020,Heymans2020}. In those analyses, the accuracy of the calibrated redshift distributions is prioritised over precision in the resulting cosmic shear measurement, using a self-organising map (SOM) to remove source galaxies from the sample whose redshift could not be accurately calibrated using spectroscopy, owing to incompleteness in the spectroscopic sample \citep{Wright2020a,Hildebrandt2020}. Since percent-level biases in the mean redshifts do not impact our analysis, we prioritise precision in the cross-correlation measurement by analysing the full KiDS-1000 source sample, calibrated with the DIR approach. We recognise that, particularly at high redshifts, the accuracy of the DIR calibration approach decreases, as the incompleteness of the spectroscopic sample increases.

For our fiducial analysis we include all galaxies with a photometric redshift $0.1 < z_{\rm B} < 1.2$. We also measure the cross-correlation with an additional high redshift sample, $1.2< z_{\rm B}<2$, with which we investigate the accuracy of the shear and redshift calibration in this sample, previously unused in any KiDS analysis. The redshift distribution for this second sample is also derived using the DIR method. The redshift distributions for the two samples are shown with the solid/dashed lines in Fig. \ref{fig:nz} \citep{Hildebrandt2020}.
 
After masking, which removes a range of contaminants including bright stars or galaxies and bad CCD pixels, the net area of KiDS-1000 is 777 deg$^2$ (shown in Fig. 1 of \citet{Kuijken2019}). The mean redshift is $0.72$ for the main sample which has an effective source galaxy density of $7.66$ gal/arcmin$^2$ \citep{Hildebrandt2020} for $(0.1<z_{\rm B}<1.2)$. The high redshift sample, $(1.2<z_{\rm B}<2)$, has a mean redshift of $1.24$ and an effective source galaxy density of $0.91$ gal/arcmin$^2$.

\subsection{CMB Lensing}
Here we briefly describe the two CMB lensing convergence maps, $\kappa_{\rm C}$, which overlap with the KiDS-1000 cosmic shear data.

\subsubsection{ACT}
The ACT CMB lensing map used in this analysis \citep{Darwish2019} is reconstructed from a sub-set of arcminute-resolution maps of the temperature and polarisation anisotropy, described in \citet{Thornton2016,Aiola2019,Choi2020}. The data used for the lensing reconstruction includes observations taken during 2014 and 2015 in the 98~GHz and 150~GHz frequency bands, and consists of two patches of sky on the celestial equator of which we use the `BN' region, which overlaps with the KiDS-N data over an area of $\approx 275$~deg$^2$. This is part of ACT's fourth data release (DR4).

CMB lensing maps are reconstructed using an estimator quadratic in the inverse-variance filtered CMB maps band-limited to some range in multipole space. Atmospheric noise and CMB map-maker convergence impose a lower limit on the angular scales that can be included from ACT CMB maps to around $\ell=500$, but this can be extended to lower multipoles if CMB maps are co-added with \textit{Planck}. An upper limit of $\ell= 3000$ is typically set so that contamination from foregrounds is small, but this can also be extended to smaller scales if mitigation techniques are used. Because low-$L$ lensing modes modulate high-$\ell$ CMB power, the lowest multipoles in the lensing field that can be reconstructed are typically lower than the lowest-$\ell$ (largest scales) used in the input CMB maps.

Potential sources of extragalactic foreground bias due to the thermal and kinetic Sunyaev-Zel’dovich  (tSZ and kSZ) effect \citep{Sunyaev1970,Sunyaev1980}, CIB and both radio and in-frared point sources, have been studied extensively \citep{vanEngelen2014,Schaan2019,Sailer2020}. For low redshift galaxy samples tSZ is expected to be the largest systematic, but with higher redshift samples the contribution from CIB is comparable \citep{Sailer2020}. For ACT the tSZ and CIB signals could bias spectra by around 10\%, which is significantly smaller than the ACT error; furthermore, the tSZ contribution, which dominates at low-$z$, is removed by the use of tSZ deprojected lensing maps, as described below. \textit{Planck} lensing is expected to have negligible foreground biases due to the lower resolution of \textit{Planck}.

\citet{Darwish2019} presented CMB lensing convergence maps reconstructed from ACT data applying two methods –- the first using only ACT data and the second using internal linear combinations (ILC) of \textit{Planck} and ACT data to remove contamination from the tSZ effect \citep{Madhavacheril2018,Madhavacheril2020}. We performed our analysis on both sets of maps to confirm that this did not impact our results. The results shown here use the tSZ-free maps, which use the range $100 < \ell < 3350$ from the CMB to reconstruct the lensing field. The lensing maps presented in \citet{Darwish2019} have been divided by the mean of the square of the mask, which we multiply back out since we do not have complete overlap on the ACT region with KiDS. Additionally, we use $511$ simulations of the ACT survey, which accurately capture the instrumental and observation noise properties of the data, to estimate our covariance matrix \citep[for more details on the estimation of the ACT convergence map and its associated simulations, see][]{Darwish2019}.

\subsubsection{\textit{Planck}}
Since ACT DR4 observations in 2014-15 only overlapped with the KiDS-N region, we additionally cross-correlate the KiDS-S data with the publicly available CMB lensing map from the \textit{Planck} 2018 data release\footnote{Planck Legacy archive: https://pla.esac.esa.int/} \citep{Planck2018Lens}. The \textit{Planck} $\kappa_{\mathrm{C}}$ map has complete overlap with the KiDS data, however we only use it on the  KiDS-S region, providing an area of $\approx 450$~deg$^2$.

As described in \citet{Planck2018Lens}, the lensing map is reconstructed from the final \textit{Planck} temperature and polarisation data, where the lensing signal has been detected to 40$\sigma$ in the auto-spectrum. The angular resolution of the data is lower than that of ACT; hence only scales in the range $(100 < \ell < 2048)$ were used in the reconstruction of the final $\kappa_{\mathrm{C}}$ map with reliable modes in the range $(8 < L < 2048)$. The data release provides two versions of this map, an optimally-weighted and an isotropically-weighted map, of which we use the former in the current paper. We have performed our full analysis on both sets of maps to ensure this choice does not impact our results. Finally, we use the 300 public simulations for covariance estimation and null tests \citep{Planck2018Lens}.

\section{Methods}
\label{sec:method}
In this section we describe our methods used to measure the cross-correlation between the galaxy lensing data and the $\kappa_{\mathrm{C}}$ maps, estimate its associated uncertainty, and model the resulting signal.

\subsection{Power Spectrum Estimation}
Starting from the KiDS-1000 shape catalogue, we construct a shear map covering the survey footprint, taking the weighted average of the galaxy ellipticities in each pixel. The KiDS-North and KiDS-South shear maps are designed to map directly onto the ACT and the \textit{Planck} convergence maps respectively,  which use different pixelisation and coordinate systems. ACT uses the Plate Carree pixelisation \citep{Calabretta2002} as described in \citet{Darwish2019}: it is a special case of equirectangular projection, in which all pixels are rectangular and have a varying area. For the KiDS-S data we first rotate the ellipticities into Galactic coordinates (the native \textit{Planck} coordinate system) and create a HEALPix map of NSIDE = 2048 \citep{healpix}, matching the resolution of the \textit{Planck} lensing maps. 
In both regions, weight maps are also created from the sum of the galaxy \textit{lens}fit weights per pixel, which we use in our estimator. 

We zero-pad all maps such that both maps in a given cross-correlation measurement cover the same area with the pixels set to zero in a given map where there is no data from that survey, even when that pixel may be non-zero in the map it is being cross-correlated with. This approach maximises the area used in our analysis, as the convergence in one pixel will be correlated with pixels outside of the map because the correlation length is greater than zero \citep{Harnois2016}.  We also avoid using apodisation, and apply the galaxy weight maps instead of the survey footprint, in order to maximise the SNR of the measurement, as in \citet{Hikage2019} \citep[See][for further discussion on power spectra analysis choices]{Harnois2016,namaster}. Instead of reconstructing a galaxy lensing convergence map, we work directly with the two shear components ($\gamma_1, \gamma_2$), equivalent to using Stokes Q and U components in a CMB polarisation analysis. We tested our pipeline, described above, on simulations with the real KiDS-1000 footprint, weights and noise in the simulated galaxy lensing shear maps and ACT noise and footprint in the simulated CMB lensing maps, ensuring that our analysis choices do not produce a biased spectrum.

From the three fields, $\{\kappa_{\mathrm{C}},\gamma_1,\gamma_2\}$, we estimate the correlation of the galaxy weak lensing E-mode with the CMB lensing convergence ($\kappa_{\mathrm{C}} \gamma_{\mathrm{E}}$) using the \texttt{NaMaster}\footnote{https://github.com/LSSTDESC/NaMaster} \citep{namaster} pseudo-$C_{\ell}$ estimator, employing the curved-sky computational settings. This method also produces an estimate of the correlation with the galaxy weak lensing B-mode as a by-product ($\kappa_{\mathrm{C}} \gamma_{\mathrm{B}}$), which is expected to be null.

We measure the cross-correlation power spectrum, $C_\ell^{\kappa_{\mathrm{C}} \gamma_{\mathrm{E}}}$ in 5 logarithmically spaced bins between $\ell=60$ and $\ell=2500$. The lower limit comes from the survey sizes while the upper limit removes modes beyond the map resolution. 

\subsection{Covariance}
The covariance of the cross-correlation power spectrum is estimated using a suite of 511 ACT CMB lensing simulations as described in \citet{Darwish2019} and the set of 300 publicly available \textit{Planck} CMB lensing simulations \citep{Planck2018Lens}. These each have a different signal and noise realisation that are consistent with the CMB lensing maps. These simulations do not correlate with the foreground galaxy weak lensing signal, and hence are only adequate for estimating the covariance matrix provided that
\begin{equation}
    (C_{\ell}^{\kappa_{\mathrm{C}}\kappa_{\mathrm{C}}} + N_{\ell}^{\kappa_{\mathrm{C}}\kappa_{\mathrm{C}}})(C_{\ell}^{\gamma_{\mathrm{E}}\gamma_{\mathrm{E}}} + N_{\ell}^{\gamma_{\mathrm{E}}\gamma_{\mathrm{E}}}) \gg (C_{\ell}^{\kappa_{\mathrm{C}}\gamma_{\mathrm{E}}})^2 \, ,
\end{equation}
for power spectra $C_{\ell}^{ij}$ and noise spectra $N_{\ell}^{ij}$, where $i,j=\{\kappa_{\rm C},\gamma_{\rm E}\}$ are the CMB lensing convergence and the galaxy lensing shear E mode respectively.
This is satisfied for the current analysis where the CMB lensing auto-spectrum and noise spectrum dominate the error budget, but will need to be revisited in future data analyses using CMB and galaxy lensing experiments. 

To construct the covariance matrix we measure the cross-power spectrum between the simulated CMB lensing maps and the real KiDS lensing maps, $C_{\ell}^{\kappa_{\mathrm{C}}^{\rm sim} \gamma_{\mathrm{E}}}$, and then estimate the covariance matrix from these measurements, as
\begin{equation}
    \mathrm{Cov}_{\ell,\ell'}^{\kappa_{\mathrm{C}}\gamma_{\mathrm{E}}} = \langle \Delta C_{\ell}^{\kappa_{\mathrm{C}}^{\rm sim} \gamma_{\mathrm{E}}} \Delta C_{\ell'}^{\kappa_{\mathrm{C}}^{\rm sim} \gamma_{\mathrm{E}}} \rangle \, ,
    \label{covariance}
\end{equation}
where the $\Delta C_{\ell}^{\kappa_{\mathrm{C}}^{\rm sim}\gamma_{\mathrm{E}}} = C_{\ell}^{\kappa_{\mathrm{C}}^{\rm sim}\gamma_{\mathrm{E}}} - \bar{C}_{\ell}^{\kappa_{\mathrm{C}}^{\rm sim}\gamma_{\mathrm{E}}}$ is computed for each simulation and $\bar{C}_{\ell}^{\kappa_{\mathrm{C}}^{\rm sim}\gamma_{\mathrm{E}}}$ is the mean. The angular bracket here refers to an average over the 511 (300) ACT (\textit{Planck}) simulations. This estimate of the covariance matrix includes the contributions from the CMB noise times the galaxy lensing noise \citep[see][for some more details]{Harnois2016}. The sample variance is not captured here which would require simulated maps of both CMB lensing and galaxy lensing with their cross-spectrum included; however this is not needed here given the current levels of noise. We also considered randomly rotating the galaxy ellipticities to better converge on the mean shape noise, but found  -- as previous studies with KiDS-450 also revealed -- that doing so almost perfectly recovers the case where the random ellipticities are not included.

\subsection{Modelling}

\subsubsection{Theory spectrum}
Our theoretical model for the unbinned cross-correlation power spectrum is given by
\begin{equation}
C_{\ell}^{ij} = \int^{\chi_*}_0 {\rm d} \chi \frac{W_{i} (\chi) W_{j}(\chi)}{f_k(\chi)^2}P\left(k=\frac{\ell + \frac{1}{2}}{\chi},\chi\right) \, , 
\label{crossspectrum}
\end{equation}
where $i$ and $j$ denote $\kappa_{\rm C}$ and $\gamma_{\rm E}$ respectively, $W_{i/j}(\chi)$ are the window functions, $\chi$ is the comoving distance, $f_k(\chi)$ is the comoving angular diameter distance, $P(k,\chi)$ is the 3D matter power spectrum at a given comoving distance and $\chi_*$ is the comoving distance to the surface of last scattering \citep{limber,Kaiser1992,Kilbinger2015}. The general form for the lensing window function is 
\begin{equation}
W_i(\chi) = \frac{3 \Omega_{\rm m} H^2_0}{2c^2} \frac{f_k(\chi)}{a(\chi)}\int^{\chi_{*}}_{\chi} {\rm d}\chi' p_i(\chi') \frac{f_k(\chi'-\chi)}{f_k(\chi')} \, ,
\label{windowfunction}
\end{equation}
where $p(\chi)$ is the comoving distance probability distribution function of the light sources and $a(\chi)$ is the scale factor. For CMB lensing, this is given by
\begin{equation}
p_{\kappa_{\rm C}}(\chi) = \delta(\chi - \chi_*) \, ,
\end{equation}
since the CMB is on a single source plane. For galaxy weak lensing, this is given by
\begin{equation}
p_{\gamma_{\rm E}}(\chi) = n(z) \frac{{\rm d}z}{{\rm d}\chi} \, ,
\end{equation}
where $n(z)$ is the redshift distribution of the galaxies, normalised to integrate to unity, and estimated in Fig. \ref{fig:nz} for the two samples used in our analysis. 

The theoretical cross-correlation power spectrum is calculated using the Core Cosmology Library  \citep[\texttt{CCL}\footnote{https://github.com/LSSTDESC/CCL},][]{CCL} which, using the default settings, estimates $P(k,\chi)$ employing the \texttt{CLASS} algorithm \citep{Blas2011}. Within \texttt{CLASS} a Boltzmann solver is used to compute the linear power spectrum and the \texttt{halofit} \citep{Smith2003,Takahashi2012} fitting function for the non-linear part of the matter power spectrum. We extend this framework to include systematic effects present in the galaxy weak lensing data,  which we describe in the following subsections.

\begin{table*}
    \centering
    \begin{tabular}{c|c|c|c|c}
          & &  & Prior & Prior  \\
          Parameter & \textit{Planck} & K-1000 & $(0.1<z_{\rm B}<1.2)$ & $(1.2<z_{\rm B}<2)$ \\
         \hline
         \hline
         $\Omega_{\rm m}$ & 0.31 & 0.25 & [0.1,0.9] & fixed \\
         $\sigma_8$ & 0.81 & 0.84 & [0.5,1.5] & fixed \\ 
         $\Omega_b$ & 0.04 & 0.04 & fixed & fixed \\
         $h$ & 0.67 & 0.77 & fixed & fixed \\
         $n_s$ & 0.97 & 0.90 & fixed & fixed \\
         \hline
         $m$ & 0.0 & 0.0 & $\mathcal{N}$(0.0,0.02) & [-2,2] \\
         $\Delta z$ & 0.0 & 0.0 & $\mathcal{N}$(0.0,0.027) & [-0.5,0.5] \\
         $A_{\mathrm{IA}}$ & \{0,0.54,1\} & {0.54} & $\mathcal{N}$(0.54,0.65) & fixed \\
         \hline
    \end{tabular}
    \caption{In this analysis we evaluate the amplitude of the cross-correlation in the context of both a \textit{Planck} and KiDS-1000 best-fit cosmology, the key parameters of which are specified in the respective columns. We use the \textit{Planck} values as our fiducial model. The final two columns contains the priors -- square brackets correspond to a top-hat prior, $\mathcal{N}$ corresponds to a Gaussian prior centred on the mean $\mu$ and the standard deviation $\sigma$}.
    \label{tab:priors}
\end{table*}

\subsubsection{Redshift Calibration Uncertainty}
Biases resulting from systematic uncertainty in the source redshift distributions have been shown to affect the cross-correlation predictions by up to 10\% in previous analyses \citep{Hand2015}. This is larger than for cosmic shear studies, and is due to the fact that photometric redshift estimation becomes most uncertain in the high-redshift tail, where the contribution to the CMB-weak lensing cross correlation signal is the strongest. Errors in the photometric redshift distribution are typically accounted for by modelling this uncertainty as a shift in the redshift distribution, 
\begin{equation}
    n(z)_{\rm biased} = n(z + \Delta z) \, ,
\end{equation}
where $n(z)_{\rm biased}$ is the observed redshift distribution. The photometric redshift bias, $\Delta z$, can take a positive or negative value, which corresponds to shifting the $n(z)$ to a higher or lower redshift about the mean of the distribution, which results in an increase or decrease of the cross-correlation power spectrum  respectively. When cross-correlating with CMB lensing we expect to be more sensitive to the high-redshift tail rather than the main body of the $n(z)$, and therefore $\Delta z$ should not be understood literally as the error on the source redshift distribution. But given our relatively low statistical precision, this is not an issue for our measurement and this parameterisation is therefore sufficient. Nevertheless, in the future we anticipate that the main $n(z)$ will be robust but the high-redshift tail will not, making this type of measurement useful for constraining the parameters which describe the tail of the distribution. Shifting the redshift distribution by $\Delta z$ changes the galaxy weak lensing window function which impacts the amplitude of the cross-correlation power spectrum. We assign a Gaussian prior to $\Delta z$, motivated by the calibration of the $n(z)$ described in \citet{Hildebrandt2020} and shown in Table \ref{tab:priors}. This prior corresponds to the  68\% confidence interval on the mean DIR redshift distribution, which is estimated from a spatial bootstrap resampling of the spectroscopic redshift calibration sample.

\subsubsection{Shear Calibration Uncertainty}
Weak lensing shear measurements are calibrated to account for biases introduced by pixel-noise, object selection and galaxy weights. The KiDS calibration correction is established through the analysis of mock image simulations that emulate the KiDS imaging \citep{Kannawadi2019}. We account for the level of uncertainty in the measured calibration correction with the inclusion of a nuisance parameter $m$, that we marginalise over in our cosmological analysis. Here the observed cross correlation, $C_{\ell,\mathrm{obs}}^{\kappa_{\rm C}\gamma_{\rm E}}$, is related to the true cross correlation, $ C_{\ell}^{\kappa_{\rm C}\gamma_{\rm E}}$ as
\begin{equation}
    C_{\ell,\mathrm{obs}}^{\kappa_{\rm C}\gamma_{\rm E}} = (1+m)C_{\ell}^{\kappa_{\rm C}\gamma_{\rm E}} \, .
\end{equation}
\citet{Kannawadi2019} determine Gaussian priors for $m (z_{\rm B})$, the calibration correction as a function of photometric redshift, out to $z_{\rm B} < 1.2$.  For our fiducial analysis with $0.1 < z_{\rm B} < 1.2$ we adopt a zero-mean Gaussian prior for $m$ with a width $\sigma_m = 0.02$, as approximated from a galaxy weighted average of $m (z_{\rm B})$, and following the level of uncertainty recommended by \citet{Kannawadi2019}. For our high-redshift analysis with $1.2<z_{\rm B} < 2$, we use a fully uninformative prior on $m$, as our aim here is to use the cross-correlation of this sample to understand galaxy weak lensing systematics.

\subsubsection{Intrinsic Alignment}
The coherent distortions of galaxy shapes resulting from gravitational lensing can be mimicked or obscurred by the intrinsic alignment of galaxy shapes. This effect is caused by the interplay of galaxy formation and evolution, and tidal torques that arise from the large-scale structure in which the galaxies are embedded
\citep[see][and references therein]{Joachimi2015,Kirk2015,Kiessling2015}.

Intrinsic alignment has two forms: the intrinsic alignment between galaxies referred to as II and correlation between the intrinsic alignment and lensing signal in cosmic shear surveys known as GI. Only the latter comes into play in a cross-correlation measurement. Indeed, the light from the CMB is lensed by the large scale structure, producing distortions that are tangentially aligned around massive foreground haloes. In the non-linear alignment model \citep{Bridle2007}, the galaxies that inhabit in the same objects will tend to have an intrinsic ellipticity that aligns radially with the centre of the mass, thereby inducing an anti-correlation with the CMB lensing signal and cancelling part of the cross-correlation signal.

This effect is captured in \texttt{CCL}, which employs the standard non-linear alignment model in which the intrinsic galaxy inertia tensor is proportional to the local tidal tensor. This enters our model as an additional term 
\begin{equation}
C_{\mathrm{obs}}^{\kappa_{\mathrm{C}}\gamma_{\mathrm{E}}} (\ell) = C^{\kappa_{\mathrm{C}}\gamma_{\mathrm{E}}} (\ell) + C^{\kappa_{\mathrm{C}} \epsilon_{\mathrm{IA}}} (\ell) \,
\end{equation}
where $C^{\kappa_{\mathrm{C}}\epsilon_{\mathrm{IA}}} (\ell)$ accounts for the cross-correlation between the intrinsic alignments of galaxies and the CMB lensing convergence. This extra term is calculated by replacing the galaxy lensing window functions in Equation \ref{crossspectrum} with
\begin{equation}
 W_{\mathrm{IA}}(\chi) = F(z)n(z)\frac{{\rm d}z}{{\rm d}\chi} \, .
\end{equation}
Here, $F(z)$ is equivalent to the product of the so-called alignment bias and the fraction of aligned galaxies in the sample, and is modelled as
\begin{equation}
    F(z) = A_{\mathrm{IA}} \frac{-C_1\rho_{c}\Omega_{\rm m}}{D(z)} \, , 
    \label{eq:IA}
\end{equation}
where $A_{\mathrm{IA}}$ is a constant that controls the amplitude of the cross-correlation, the constant $C_1$ absorbs the dimensions of the normalisation such that we set $C_1\rho_c = 0.0134$ and $D(z)$ is the growth factor \citep{Joachimi2011}. Given the relatively low SNR of our measurement, we ignore any redshift or luminosity dependence and simply fit for the $A_{\mathrm{IA}}$ term. Observational studies of intrinsic alignment have been limited to bright samples \citep[e.g.,][]{Johnston2019}. To be able to relate these measurements to the fainter and high redshift galaxy samples used in analyses like this one requires the correct modelling to account for the difference in the blue/red galaxy fraction and also the satellite galaxy fraction. We therefore adopt a wide prior motivated by the model presented in \citet{Fortuna2020}; a detailed discussion is presented in Appendix \ref{app}. For the cosmological analysis we use a Gaussian prior centred on $A_{\rm{IA}}=0.54$ with a width of $\sigma_{A_{\rm IA}}=0.65$. This prior captures the upper and lower values of $A_{\rm IA}$ from the KiDS-1000 cosmic shear analysis and the cases presented in \citet{Fortuna2020}.

\subsubsection{Binned measurement}

Once the previous systematic effects have been ingested in our model, the next step is to bin the model in the same way as the data. From our theoretical cross-spectrum $C_\ell$ we compute the binned model, which accounts for the mode coupling matrix, $M$, and the binning matrix, $B$, both computed by \texttt{NaMaster} \citep[equations 9-19 in ][]{namaster}.
$M$ takes into account the fact that a spherical harmonic transform of a (masked) non-periodic map couples different Fourier modes, while $B$ provides a mapping from a finely sampled $\ell$ measurement to our selected broad bin space. 

Our pipeline was tested using simulations to ensure none of our choices could produce a biased spectrum. From the set of convergence realisations used to create the ACT lensing simulations described in \citet{Darwish2019} we produced correlated realisations of the galaxy lensing shear fields. Using our pipeline to measure the cross-correlation spectra between the suite of simulated ACT lensing maps and simulated shear fields we found we could reproduce the cross-correlation spectrum from our input cosmology to within $0.2 \sigma$. Within this framework we also checked that choices regarding the galaxy weak lensing mask did not produce a biased spectrum.

\subsection{Estimating parameters}
Given our measured cross-correlation power spectrum $C_{\mathrm{obs}}^{\kappa_{\mathrm{C}}\gamma_{\mathrm{E}}} (\ell)$, an inverse covariance matrix $\mathrm{Cov}^{-1}_{\ell,\ell'}$ and a model $C^{\kappa_{\mathrm{C}}\gamma_{\mathrm{E}}} (\ell)$ evaluated at a set of cosmological and nuisance parameters $p$, we construct a Gaussian likelihood as:
\begin{equation}
\begin{split}
\ln &\mathcal{L} [C_{\mathrm{obs}}^{\kappa_{\mathrm{C}}\gamma_{\mathrm{E}}} (\ell) | C^{\kappa_{\mathrm{C}}\gamma_{\mathrm{E}}} (\ell,p)] = \\
& -\frac{1}{2} [C_{\mathrm{obs}}^{\kappa_{\mathrm{C}}\gamma_{\mathrm{E}}} (\ell)-C^{\kappa_{\mathrm{C}}\gamma_{\mathrm{E}}} (\ell,p)]^T\mathrm{Cov}^{-1}_{\ell,\ell'}[C_{\mathrm{obs}}^{\kappa_{\mathrm{C}}\gamma_{\mathrm{E}}} (\ell')-C^{\kappa_{\mathrm{C}}\gamma_{\mathrm{E}}} (\ell',p)] \, ,
\end{split}
\end{equation}
to within an overall additive normalisation factor. We apply a correction of $\alpha=(N_{\rm sim}-N_{\rm bin}-2)/(N_{\rm sim}-1)=0.98$ to the inverse covariance matrix to address the biases that result from matrix inversion when estimating the covariance matrix from simulations \citep{Kaufman1967,Hartlap2007}.

The model has two free cosmological parameters ($\Omega_{\rm m}$ and $\sigma_8$) and three systematics parameters ($m$, $\Delta z$ and $A_{\mathrm{IA}}$), with prior values shown in Table \ref{tab:priors}. The other cosmological parameters are held fixed to the best fit values found in  \citet{Planck2018}. 

We also consider a simpler case where no cosmological parameters are varied, and instead the theoretical spectrum for the fiducial cosmological model is scaled by a free amplitude, $A$, as in e.g. \citet{Hand2015}. For this case, we scale predictions that assume best fit cosmologies derived from either the \textit{Planck} 2018 primary CMB anisotropies or KiDS-1000 cosmic shear.
 
We sample the five-dimensional likelihood with the affine-invariant MCMC sampling  code {\sc emcee}\footnote{http://dfm.io/emcee/current/} \citep{emcee}. The posterior distribution of the model parameters is given by
\begin{equation}
P(C^{\kappa_{\mathrm{C}}\gamma_{\mathrm{E}}} (\ell,p)|C_{\mathrm{obs}}^{\kappa_{\mathrm{C}}\gamma_{\mathrm{E}}} (\ell)) \propto \mathcal{L} [C_{\mathrm{obs}}^{\kappa_{\mathrm{C}}\gamma_{\mathrm{E}}} (\ell) | C^{\kappa_{\mathrm{C}}\gamma_{\mathrm{E}}} (\ell,p)] P_{\mathrm{prior}}(\pmb{p}) \, ,
\end{equation}
where priors on the model parameters are given by $ P_{\mathrm{prior}}(\pmb{p})$.

\section{Results}
\label{sec:results}

\subsection{ACT/\textit{Planck}$\times$KiDS Cross-Spectrum}
\begin{figure}
\centering
\includegraphics[width=\columnwidth]{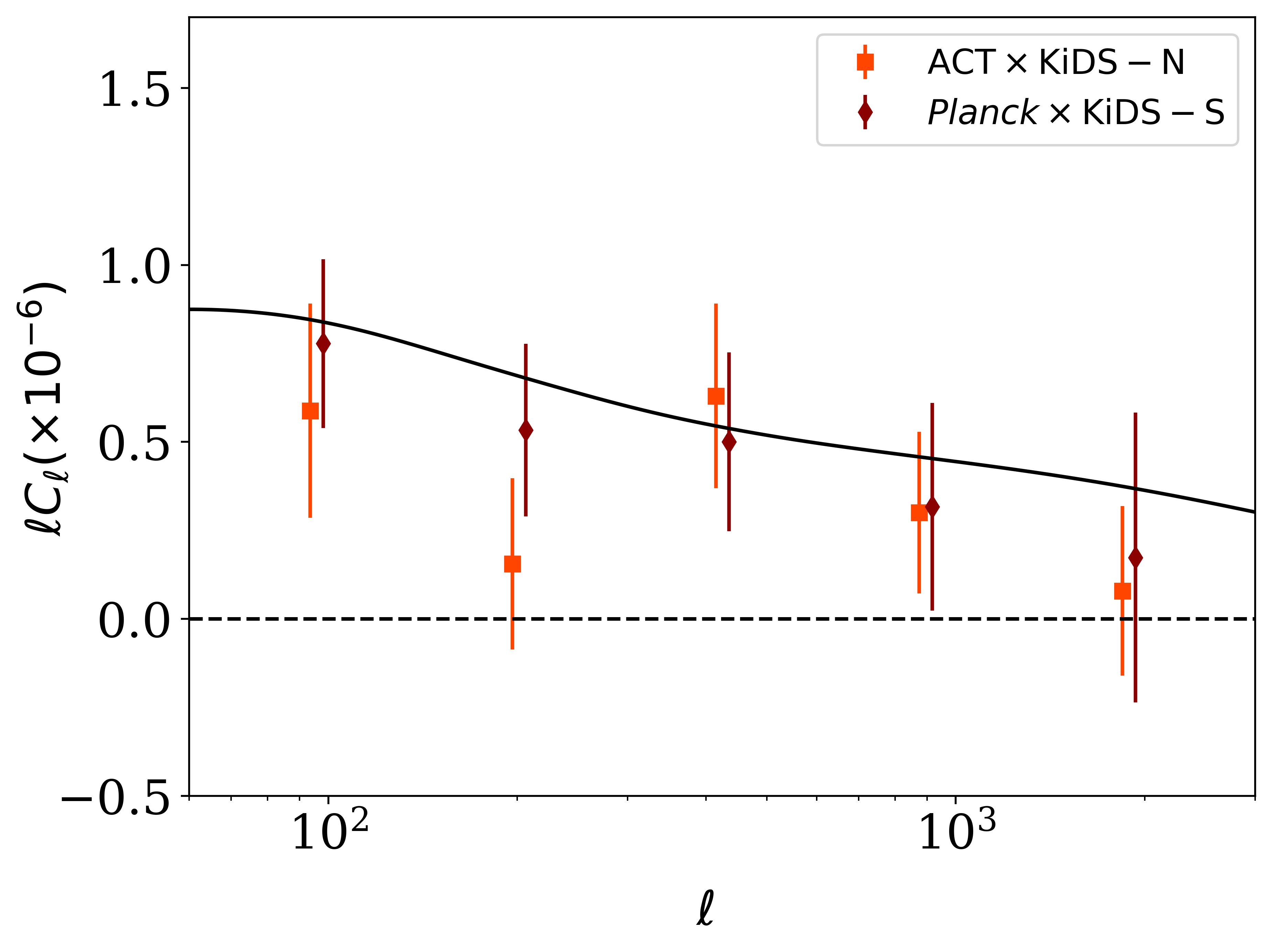}
\caption{The \textit{Planck} best-fit model, from primary CMB anisotropies, of the cross-correlation E-mode is plotted as the solid black line. We detect non-zero correlation power at $3.1\sigma/4.3\sigma$ significance for ACT/\textit{Planck}$\times$KiDS respectively. }
\label{fig:Emode}
\end{figure}

Our cross-spectrum measurements, using galaxies in the range $0.1<z_{\rm B}<1.2$, are shown in Fig. \ref{fig:Emode} where the solid black line corresponds to the best fit values found in \citet{Planck2018} as given in Table \ref{tab:priors}. The ACT/\textit{Planck}$\times$KiDS cross-correlation signal is detected with a $3.1/4.3 \sigma$ significance, compared to a hypothesis with zero signal. The significance is computed with a likelihood ratio test as $\sqrt{(\chi^2_{\rm null} - \chi^2_{\rm best\,fit})} $. In this analysis, even though \textit{Planck}$\times$KiDS-S has a larger area, the ACT noise is lower, particularly at high-$\ell$ and as a result the size of the error bars from these measurements are comparable. Combining both measurements, we obtain a combined significance of $5.3 \sigma$. This exploits the independence of both measurements, since they involve different areas of sky.

\subsection{Null Tests}
\begin{figure}
\centering
\includegraphics[width=\columnwidth]{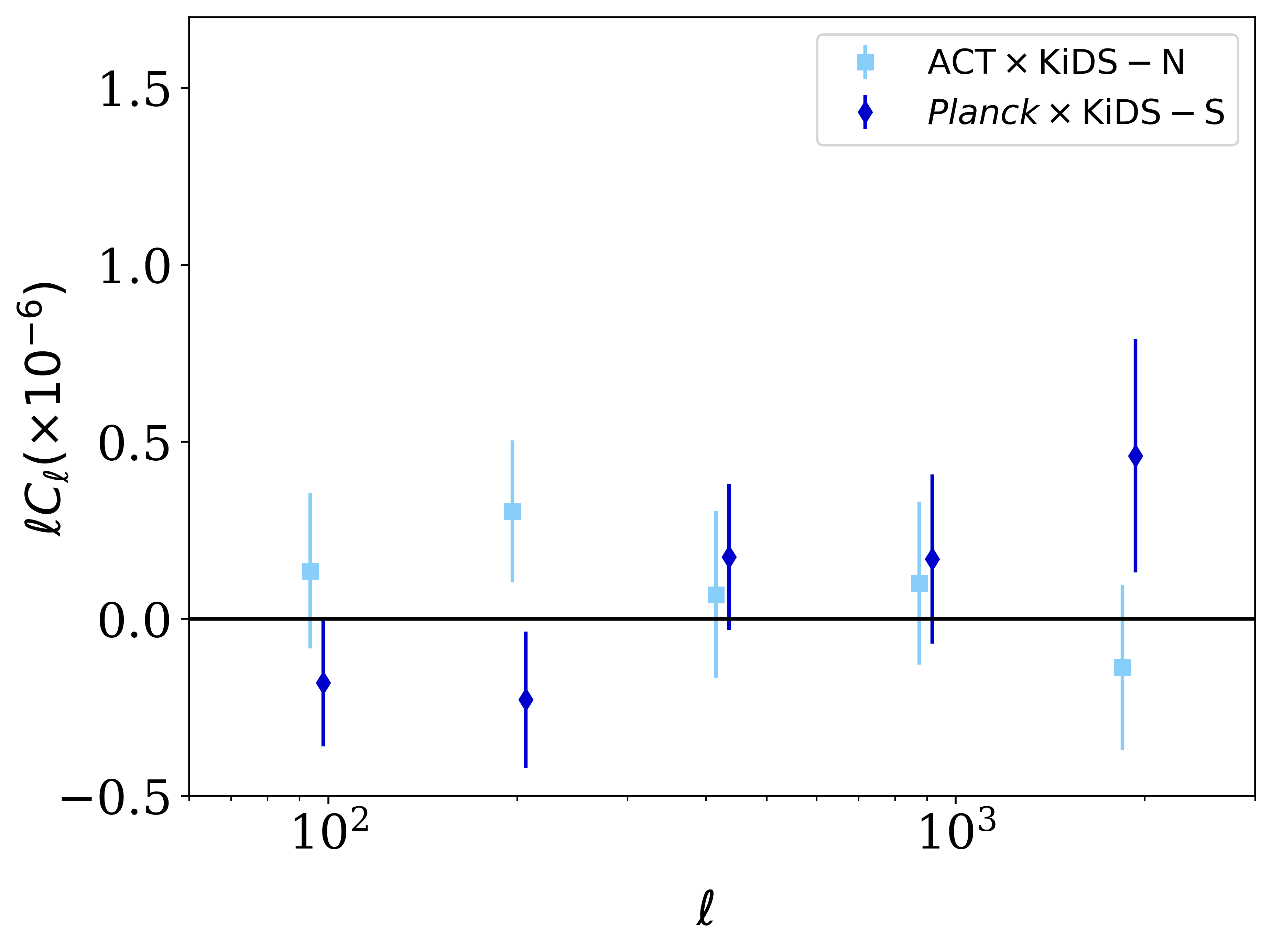}
\caption{The B-mode null tests, $C_{\ell}^{\kappa_{\mathrm{C}}\gamma_{\mathrm{B}}}$, are consistent with zero with a $p$-value of $0.8/0.9$ respectively using the full covariance.}
\label{fig:Bmode}
\end{figure}

We use the B-mode cross-correlation to assess whether there has been any E to B mode leakage due to masking, and to investigate any B-mode component that could have been produced from residual systematic effects in the shape measurement process. This measurement of $C_\ell^{\kappa_{\mathrm{C}} \gamma_{\mathrm{B}}}$ is shown in Fig. \ref{fig:Bmode} and is consistent with zero signal for both measurements. Compared to null, this measurement has a $p$-value of $0.6/0.3$ and a reduced chi-squared of $0.7/1.2$ for ACT/\textit{Planck}$\times$KiDS. 

\begin{figure}
	\includegraphics[width=\columnwidth]{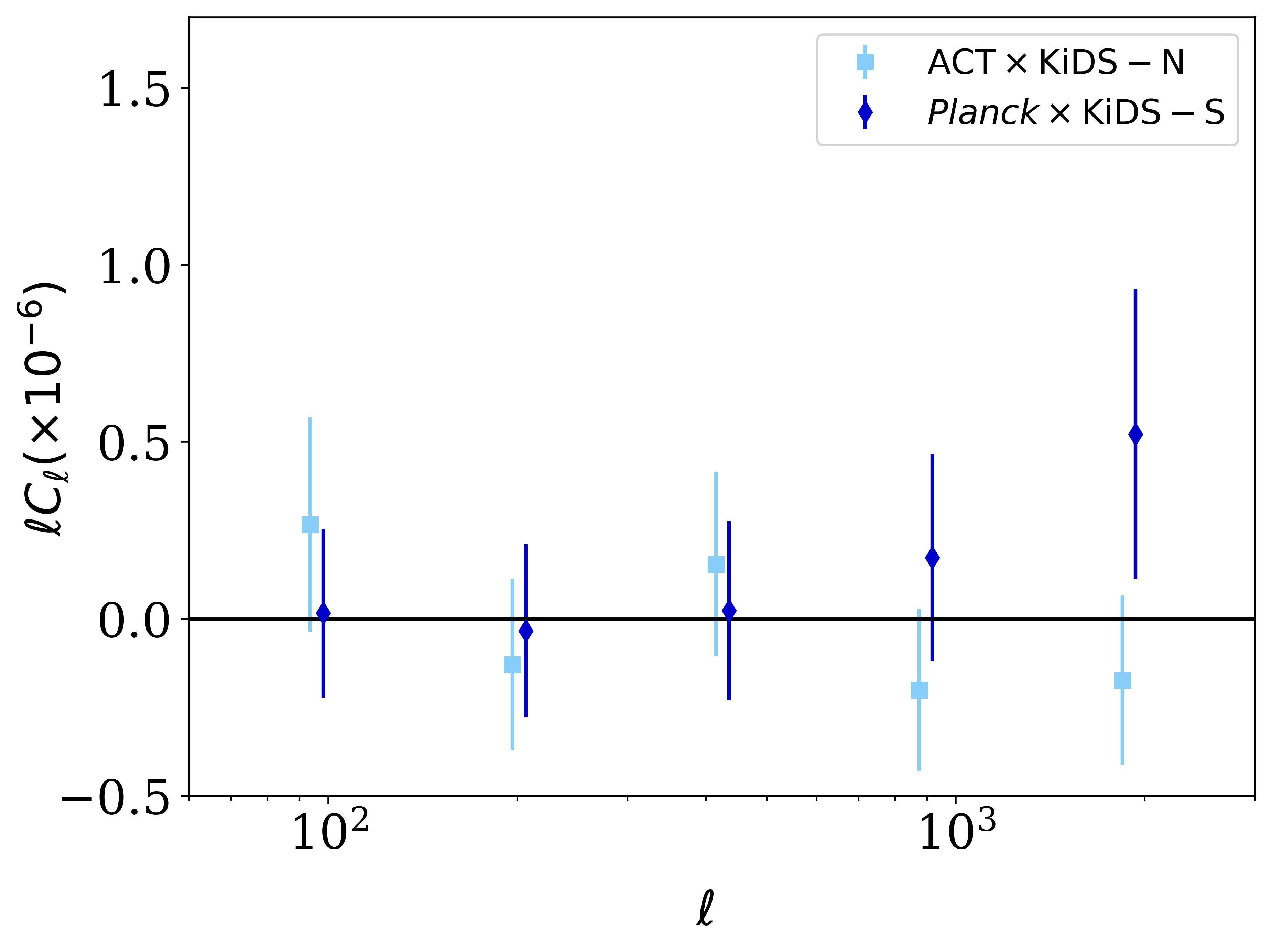}
    \caption{Random null test, in which the galaxies have been randomly rotated before measuring the cross-correlation signal. We find that this cross-spectrum is consistent with zero as expected, with $p$-values of 0.8 and 0.9 for ACT/\textit{Planck}$\times$KiDS respectively }
    \label{fig:random}
\end{figure}

We also perform a null test in which we randomly rotate the galaxies and remeasure the cross-correlation on the random map. Fig. \ref{fig:random} shows the result of this randoms test, compared to the null hypothesis. The test passes with $p$-values of $0.8/0.9$ and a reduced chi-squared of $0.4/0.3$, respectively. Although most systematic effects in the data should not produce B-modes in a cross-correlation measurement, both of these tests provide a useful check of our measurement pipeline. 

\subsection{Cosmology: Amplitude Fit}
We first perform the single-parameter fit, constraining the amplitude $A$ of the cross-spectrum relative to a fiducial cosmology (and detailed in the second column of Table \ref{tab:priors}). We ignore the uncertainty on redshift and shape calibration for the moment. If the data are consistent with the model, we expect to find $A=1$ within the uncertainties of our measurement. We consider an intrinsic alignment amplitude of $A_{\rm IA}=0.54$ as our fiducial value in Eq. (\ref{eq:IA}), as motivated by \citet{Fortuna2020}, and find $A^{\rm N}_{\it Planck}=0.56 \pm 0.19$ for the northern region from ACT$\times$KiDS-N, and $A^{\rm S}_{Planck}=0.84\pm0.19$ in the south from \textit{Planck}$\times$KiDS-S. The amplitude fits from these to patches differ by $\sim 1.5\sigma$ which is consistent with the previous results that use data from different parts of the sky such as the CFHTLenS/RCSLenS analysis \citep{Harnois2016}. Performing a combined fit we find $A_{\it Planck}=0.69 \pm 0.14$. Including zero contribution from intrinsic alignment, $A_{\rm IA}=0$, we find a lower value as expected of $A_{\it Planck}^{{\rm noIA}}=0.64 \pm 0.13$ which is consistent with earlier works that did not include the impact of intrinsic alignment \citep{Hand2015,Kirk2016,Harnois2016}. Considering an upper value of $A_{\rm IA}=1$, we find $A_{\it Planck}^{{\rm IA}=1}=0.73 \pm 0.14$. The three values of $A_{\rm IA}$ considered span the expected range from data and simulations. Even with this higher value for the intrinsic alignment we still find a cross-spectrum amplitude that is lower than predicted by a \textit{Planck} cosmology, by nearly $2\sigma$.

We reproduced the analysis in the KiDS-South region using the isotropically weighted \textit{Planck} lensing convergence maps. We find consistent results, with a shift in the cross-correlation amplitude of less than $0.4\sigma$, accompanied by a factor $\sim2$ reduction in the signal-to-noise ratio owing to the sub-optimal noise level in this map.

\begin{figure}
    \centering
    \includegraphics[width=\columnwidth]{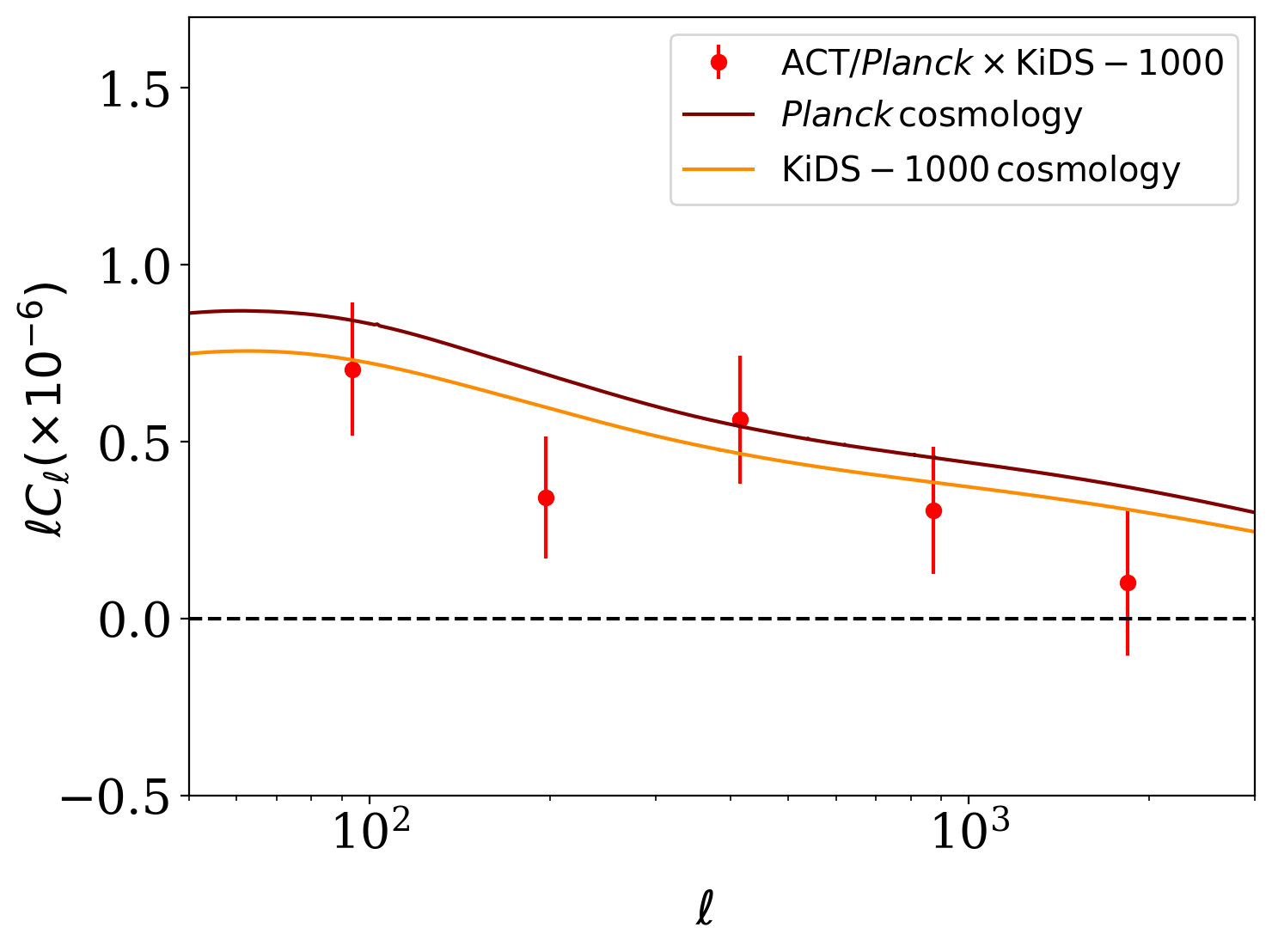}
    \caption{The theoretical cross correlation power spectrum assuming a \textit{Planck} or KiDS-1000 cosmology, compared to the combined-survey results estimated by inverse-variance weighting the ACT$\times$KiDS-N and \textit{Planck}$\times$KiDS-S (shown here in red). }
    \label{fig:KV-450vPlanck}
\end{figure}

Given the current tension between galaxy weak lensing and CMB experiments \citep[a $2-3\sigma$ tension in the $\Omega_m-\sigma_8$ plane with \textit{Planck} is reported in][]{Heymans2020,Joudaki2020,Asgari2020a,Abbott2018,Hikage2019} it is interesting to examine whether our measurement prefers a \textit{Planck} or a KiDS-1000 cosmology \citep{Asgari2020} with this single-parameter approach. Fig. \ref{fig:KV-450vPlanck} compares the combined cross-correlation measurement with \textit{Planck} and KiDS-1000 cosmologies (in maroon and orange, respectively). It is clear that given the size of our statistical error we are unable to robustly distinguish between the two cases, particularly since the choice in intrinsic alignment amplitude has a comparable impact. Nevertheless, comparing to a KiDS-1000 cosmology -- specified in the third column of Table \ref{tab:priors} -- we find $A^{\rm N}_{\mathrm{KiDS-1000}}=0.62 \pm 0.20$ and $A^{\rm S}_{\mathrm{KiDS-1000}}=0.91 \pm 0.21$ and combined fit of $A_{\mathrm{KiDS-1000}}=0.75 \pm 0.15$.

It is worth mentioning that differences with previous measurements could also be explained in part by slight differences in the baseline cosmology, although these induce variations on $A$ that are much smaller than the statistical error.

\subsection{Cosmology: $\Lambda$CDM Parameter Fits}

Here we perform the 5-parameter joint fit to the ACT$\times$KiDS-N and \textit{Planck}$\times$KiDS-S cross-correlation spectra. Our constraints on the galaxy weak lensing systematic parameters ($m$, $\Delta z$ and $A_{\mathrm{IA}}$) are dominated by the imposed priors, and given the relatively low SNR of our measurement we would not expect to be able to constrain both cosmological and galaxy weak lensing systematics parameters simultaneously. 

\begin{figure}
	\includegraphics[width=\columnwidth]{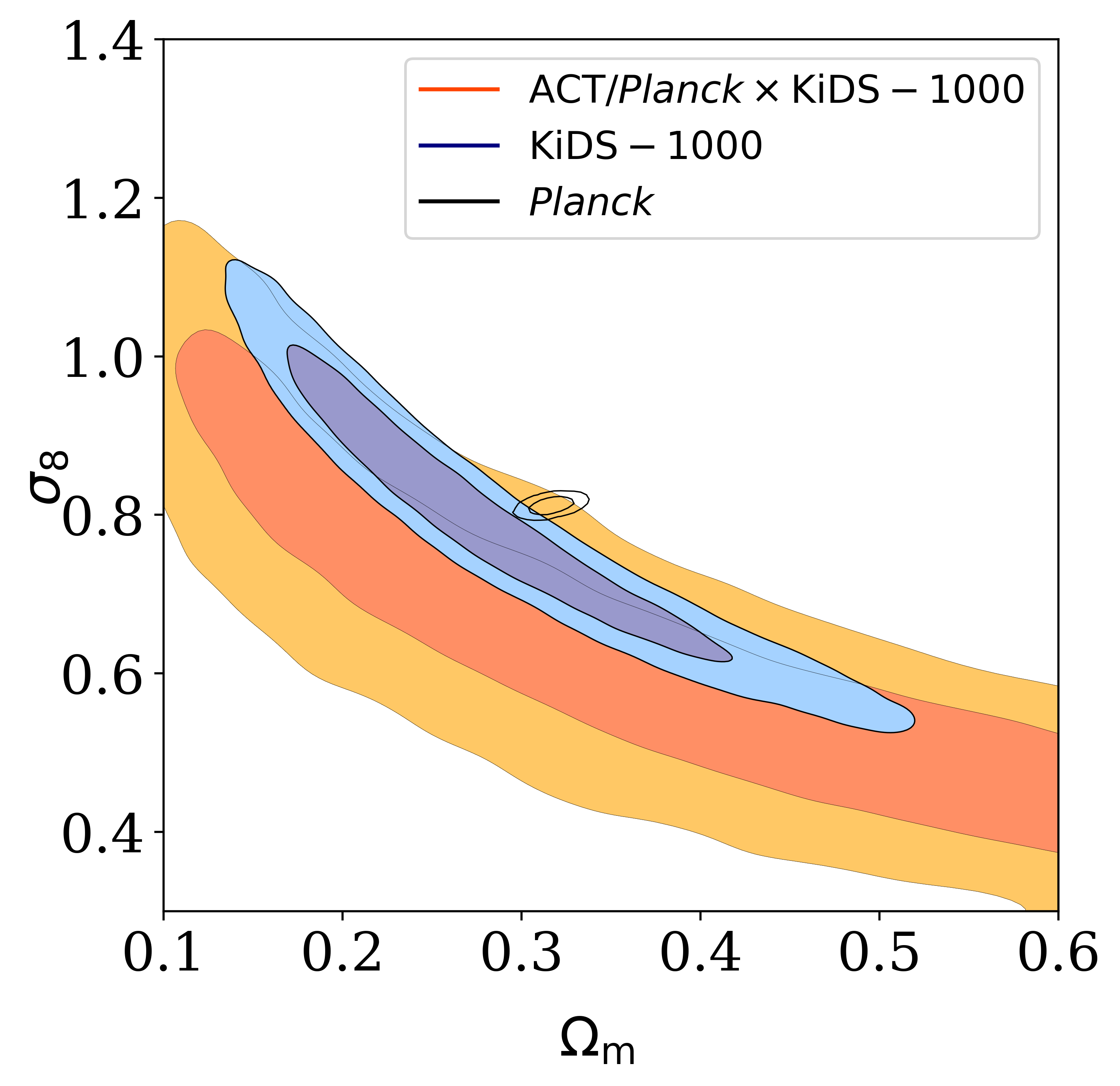}
	\includegraphics[width=\columnwidth]{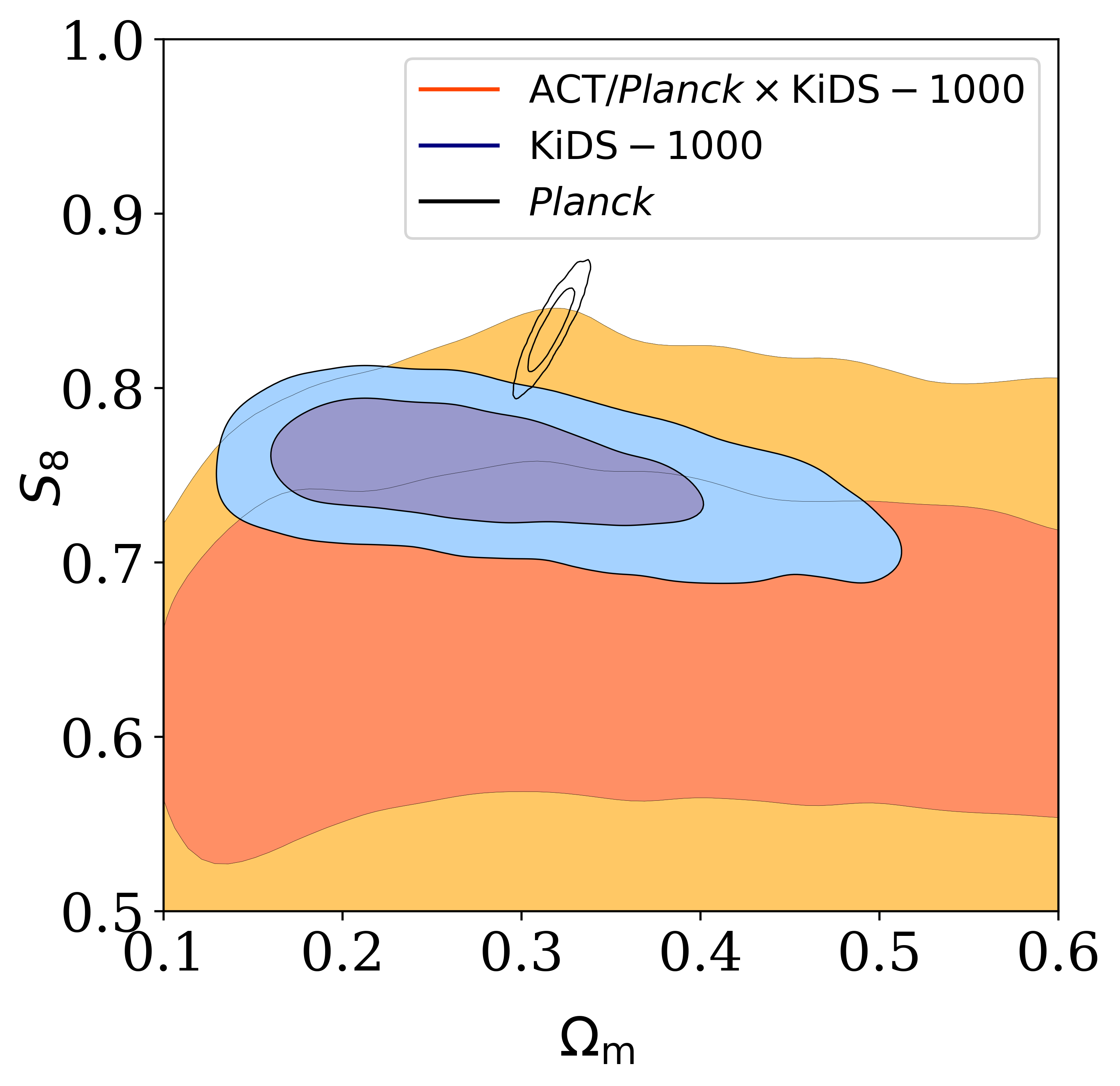}
    \caption{(Top) the posterior distributions for the amplitude of matter fluctuations, $\sigma_8$, and the matter density, $\Omega_{\rm m}$, compared to results from the KiDS-1000 cosmic shear results of \citet{Asgari2020}. (Bottom) Alternative parameterisation of these two parameters, with $S_8=\sigma_8 \sqrt{\Omega_{\rm m}/0.3}$.}
    \label{fig:S8}
\end{figure}

Our constraints are presented in Fig. \ref{fig:S8} and compared to the KiDS-1000 cosmic shear constraints \citep{Asgari2020}. These two cosmological parameters shown are strongly correlated, as seen in previous similar analyses. Weak gravitational lensing is mainly sensitive to the parameter combination $S_8=\sigma_8 \sqrt{\Omega_{\rm m}/0.3}$, which we also show in Fig. \ref{fig:S8}. We find $S_8 = 0.64\pm0.08$ at 68\% confidence. This is consistent with the KiDS-1000 cosmic shear analysis of \citet{Asgari2020}, which found $S_8=0.76\pm0.02$.
The uncertainties (in this work) are larger but consistent with the KiDS-1000 data, and we observe a slight difference in the degeneracy direction, which could be used to break the degeneracy if the noise in the $\kappa_{\mathrm{C}}$ maps was lower.

We test that if $h$ and $n_s$ are allowed to vary, the chains for these parameters reach the prior boundaries -- when using the same priors imposed on the KiDS analysis of \citet{Hildebrandt2020} using just galaxy lensing. This reflects the fact that $h$ and $n_s$ are not well constrained by our data: hence in our analysis presented here they are fixed. We recover the same best-fit value for $S_8$ if they are not fixed and we find that the degeneracy direction is not impacted.

\subsection{Redshift calibration of a high-$z$ sample}
As the mean redshift of the source galaxies grows, it becomes increasingly difficult to calibrate both the redshift and shear estimates of the galaxies. For this reason galaxies with a photometric redshift greater than 1.2 have not been included in the nominal KiDS-1000 cosmic shear analyses \citep{Asgari2020,Heymans2020}. Cross-correlations with CMB lensing has been identified as a potential calibration tool at such high redshifts \citep{Das2013,Harnois2017,Schaan2020}, and we investigate this here by measuring the $C_{\ell}^{\kappa_{\mathrm{C}} \gamma_{\mathrm{E}}}$ signal with galaxies from the KiDS high redshift sample, defined as $1.2 < z_{\rm B} < 2$. We emphasise that this calibration method for the KiDS-1000 redshift distribution is expected to be noisy given the current noise in the CMB data \citep[but this situation will soon change, as described in e.g.,][for a Rubin Observatory$\times$CMB-S4 forecast]{Schaan2020}, and it will likely be outperformed by other $n(z)$ calibration methods in the upcoming KiDS Legacy data release such as the SOM or the cross-correlation redshifts presented in \citet{Wright2020a,Hildebrandt2020,vandenbusch2020}. It nevertheless constitutes an essential validation test and offers an opportunity to further increase the strength of the $C_{\ell}^{\kappa_{\mathrm{C}} \gamma_{\mathrm{E}}}$  detection.

\begin{figure}
	\includegraphics[width=\columnwidth]{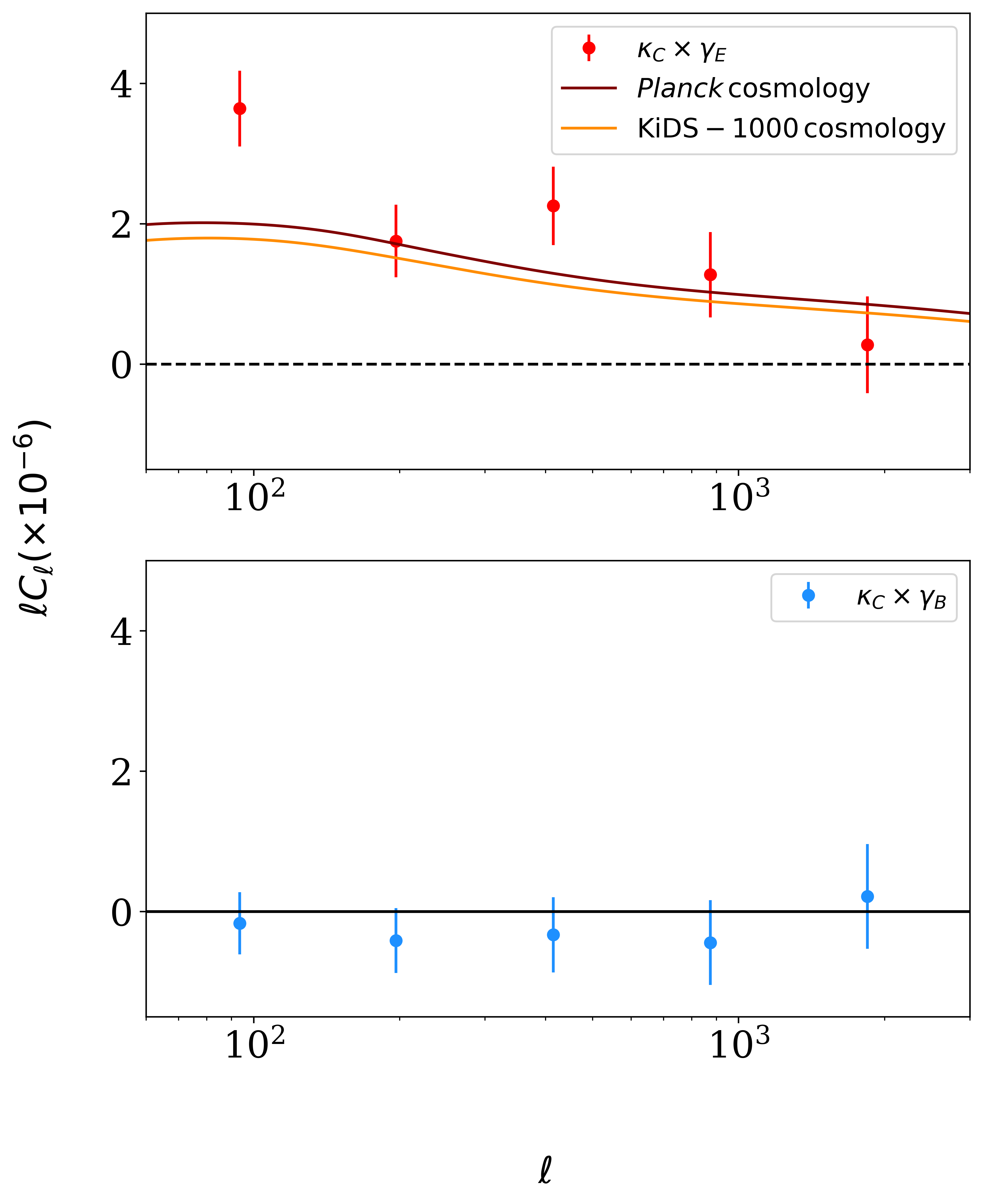}
    \caption{Combined measurement of ACT/\textit{Planck}$\times$KiDS-1000, using galaxies with a photometric redshift $1.2<z_{\rm B}<2$. Solid black line is the \textit{Planck} 2018 best fit using the redshift distribution shown in Fig. \ref{fig:nz}. The B-modes, shown in the lower panel, are consistent with zero, while the combined E-modes are detected at $7\sigma$.}
    \label{fig:bin6}
\end{figure}

Using the same tools as for the main analysis, we measure the `high-$z$' ACT/\textit{Planck}$\times$KiDS-1000 signal at a significance of $4.1 \sigma /5.7 \sigma$, with a combined significance of $7 \sigma$.With this higher redshift galaxy sample the overlap between the galaxy lensing and CMB lensing kernels increase. We therefore detect the cross-correlation with a greater significance even though our sample contains fewer galaxies compared to the lower redshift analysis. The combined E-mode measurement is shown in the top panel of Fig. \ref{fig:bin6}, over-plotted with the \textit{Planck} 2018 best fit prediction assuming the redshift distribution shown in Fig. \ref{fig:nz}. The B-mode null test is consistent with zero, which indicates the E-mode is clean of significant parity-violating effects at the same level as the lower redshift measurement.

\begin{figure}
    \centering
    \includegraphics[width=\columnwidth]{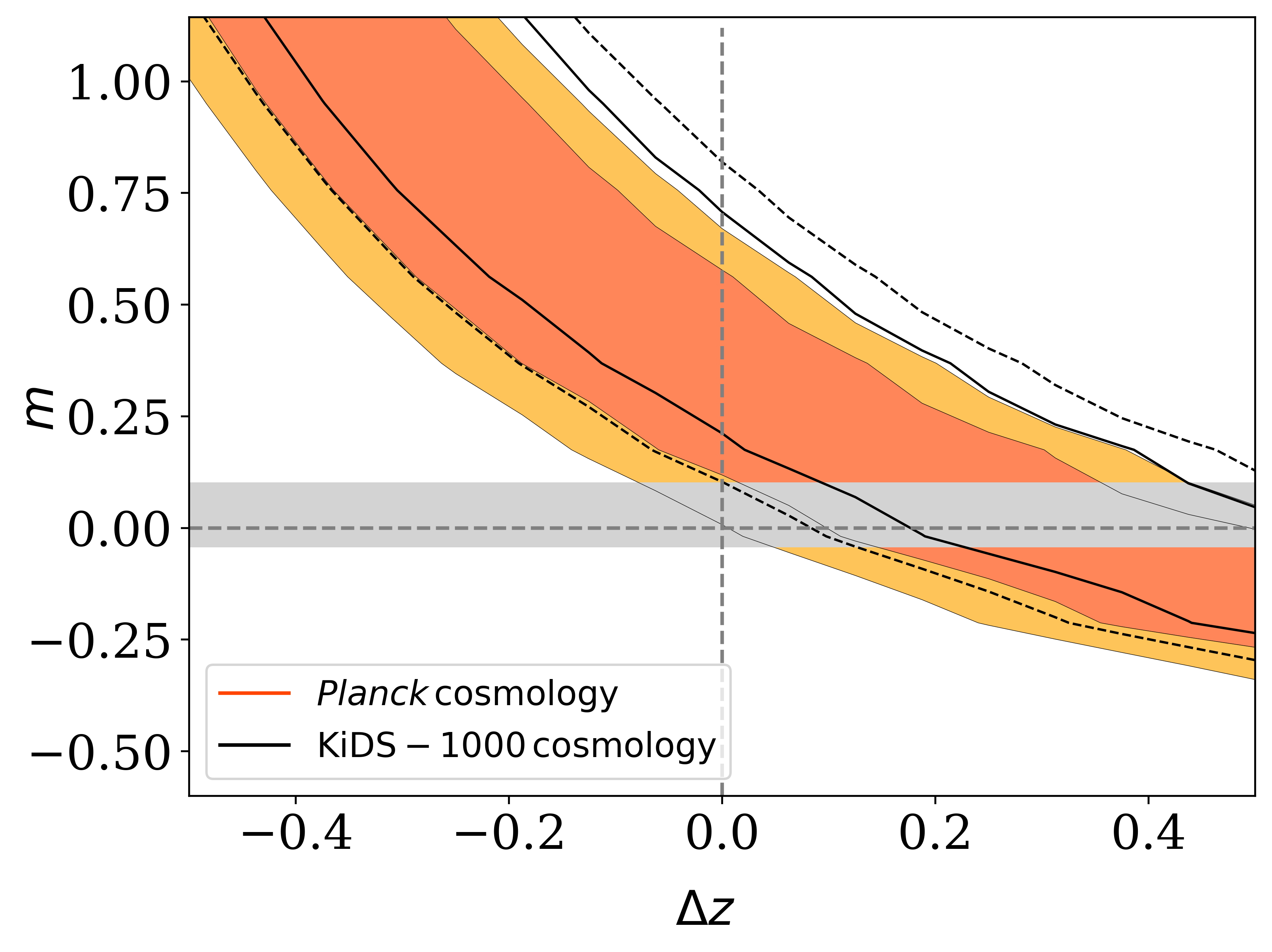}
    \caption{Joint constraint, from the high-redshift KiDS sample $(1.2<z_{\rm B}<2)$, on the two systematic parameters, $m$ and $\Delta z$, we wish to calibrate, assuming a \textit{Planck} 2018 cosmology (KiDS-1000 cosmology is shown as the black  contours). The grey band corresponds to the range of values for $m$ based on our current expectations from image simulations. }
    \label{fig:bin6_sys}
\end{figure}

We turn now to parameters of the model. We follow the same modelling methods described as for the main cosmological inference analysis, but here we fix the cosmological parameters to a \textit{Planck} 2018 cosmology and constrain $\Delta z$ and $m$. It is anticipated that intrinsic alignment is highly suppressed at such high redshifts \citep{Chisari2017,Bhowmick2020}, hence we set the intrinsic alignment amplitude $A_{\mathrm{IA}}=0$.

\citet{Harnois2017} performed a similar cross-correlation analysis, using earlier KiDS-450 and \textit{Planck} lensing data and showed a strong degeneracy between $m$ and $\Delta z$, suggesting that informative priors on at least one of the two calibration parameters would be required for this method to be competitive. We begin by setting wide, flat priors for $m$ and $\Delta z$, thus leaving them unconstrained. The posterior distribution, presented in Fig. \ref{fig:bin6_sys}, shows that the combination of $m = 0$ and $\Delta z = 0$ is excluded at nearly $2 \sigma$ for either the \textit{Planck} 2018 or KiDS-1000 cosmology; that is, one or both of these corrections are preferred to be non-zero. No shear calibration has yet been applied for this sample of galaxies and therefore we do not expect it to be exactly zero. Also indicated in Fig. \ref{fig:bin6_sys} we highlight the range of values for $m$ based on our current expectations from image simulations (shown here as the grey band).

Some offset in mean redshift distribution, $\Delta z$, is expected from the current implementation of the DIR method. Indeed, the spectroscopic redshift calibration sample is incomplete at such high redshifts, and many galaxies with photo-$z>1.5$ do not show strong spectral features in the optical bands and therefore end up being assigned no spectroscopic redshift. As a result, our selection contains many real KiDS galaxies in a high-redshift tail which are not captured by the current implementation of the DIR method, and hence are not accounted for in the $n(z)$. 

The range of values allowed for $m$ is much larger than the few percent accuracy reached with dedicated image simulations such as those described in \citet{Kannawadi2019}. We anticipate that the multiplicative bias parameter could be constrained to certainly better than $0.1$ for this tomographic bin, (the five other redshift bins in KiDS-1000 have an uncertainty under 0.02), in which case the cross-correlation measurement would mostly constrain the redshift error. Given that our expectation from image simulations is that $m$ will be close to zero for this sample, the preference for a positive shift in our measurement of $\Delta z$ points in the expected direction and indicates that the mean redshift of the $n(z)$ has been underestimated by around $\Delta z = 0.2$, due to the missing contribution from the high-$z$ tail. A non-zero value of $\Delta z$ could be capturing modifications to the redshift distribution beyond the mean, such as changes in the width and shape.

We also note that the DIR is not the only method to estimate the $n(z)$. In particular, the self-organising map method described in \citet{Wright2020} also has a high potential to constrain $\Delta z$ at high-$z$. After we achieve tight priors on both $\Delta z$ and $m$, the strong signal detected in this tomographic bin will likely dominate the cosmological constraining power. When including the measurements of the CMB lensing/galaxy weak lensing cross-correlation from the high redshift source sample, the detection significance increases from 5.3$\sigma$ to $7.7\sigma$, the highest achieved to date. This estimate of the significance accounts for the correlation between the low and high redshift measurements.

We finally caution that the treatment of intrinsic alignment is particularly uncertain here, since studies of both simulations and data have been limited to lower redshift samples. \citet{Harnois2017} estimated that for galaxies in the KiDS-450 sample, contamination from intrinsic alignment of red galaxies was of order $\sim$1\% and $\sim$3-4\% for blue galaxies, however no alignment of blue galaxies has been detected in data. At the redshifts included in the high-$z$ cross-correlation, the fraction of red galaxies is small such that the contamination from intrinsic alignment is expected to be consistent with zero \citep{Harnois2017}. This will require further investigation for future surveys, likely involving sufficiently deep hydrodynamical simulations and observations of high-redshift spectroscopic galaxies with accurate shapes, but in the mean time this justifies our choice to ignore intrinsic alignment in this section.

\section{Summary}
\label{discussion}
In this analysis we measured the cross-spectrum between the ACT and \textit{Planck} CMB lensing data and the KiDS-1000 weak lensing catalogues, selecting galaxies in the photometric redshift range of 0.1 to 1.2. The combined cross-correlation is detected with a significance of $5.3 \sigma$, and is free of B-modes and other measurable systematics -- as verified with null tests on rotated galaxy shapes.

From the measured cross-spectrum we first fitted an amplitude to re-scale a binned theoretical model employing best-fit cosmological parameters from both the \textit{Planck} 2018 primary CMB measurements and KiDS-1000; we find $A_{Planck}= 0.69 \pm 0.14$, and $A_{\mathrm{KiDS-1000}}=0.75 \pm 0.15$. These amplitudes are consistent with previous cross-correlation analyses using other data sets.

We next performed a five-dimensional cosmological inference analysis, sampling the likelihood in $\Omega_{\rm m}$ and $\sigma_8$, and marginalising over the uncertainty in the mean of the redshift distribution, shape calibration bias and the intrinsic alignment amplitude, and found $S_8\equiv \sigma_8 \sqrt{\Omega_{\rm m}/0.3} =0.64\pm0.08$, consistent with the KiDS-1000 shear analysis.

Finally we exploited the cross-correlation measurement as a tool for calibrating the redshift distributions out to a photometric redshift of 2. We measured the cross-correlation between ACT/\textit{Planck} with the KiDS-1000 high redshift galaxy sample at a significance of $7 \sigma$. We performed a joint fit for the redshift distribution and shape bias systematics parameters as a calibration exercise, fixing cosmological parameters to the \textit{Planck} 2018 primary CMB anisotropies best-fit cosmology. We showed that the degenerate combination of $m$ and $\Delta z$ can be weakly constrained, excluding the no-bias hypothesis to around $2\sigma$. We caution that a better understanding of the galaxy weak lensing systematics at these redshifts is required -- notably the effect of intrinsic alignment which was ignored in this calibration -- as we see that this cross-correlation measurement constrains a nearly fully degenerate combination of these three systematic effects. Nevertheless, this provides an independent consistency check of standard calibration methods, a redundancy that will be critical in future surveys. The next CMB lensing data from Advanced ACT will have full coverage of the KiDS region and have a lower noise level, and so there is potential for improved constraints from these data sets in the near future. While achieving greater depths is challenging, the cosmological outcome is rich: including the high redshift sample increases our detection significance, which reaches $7.7\sigma$, the highest to-date for any CMB lensing/galaxy lensing cross-correlation measurement.

\begin{acknowledgements}
The figures in this work were created with \texttt{matplotlib} (Hunter2007), making use of the \texttt{numpy} (Oliphant2006), \texttt{scipy} (Jones2001), \texttt{astropy} \citep{astropy} and \texttt{pixell} \footnote{https://github.com/simonsobs/pixell} software packages.\\
DA acknowledges support from the Beecroft Trust, and from the Science and Technology Facilities Council through an Ernest Rutherford Fellowship, grant reference ST/P004474.
JHD acknowledges support from an STFC Ernest Rutherford Fellowship (project reference ST/S004858/1).
HHi acknowledges the European Research Council (ERC) under grant agreement No.~770935 and support from the Deutsche Forschungsgemeinschaft Heisenberg grant Hi 1495/5-1.
OD, BDS, FQ and TN acknowledge support from an Isaac Newton Trust Early Career Grant and from the European Research Council (ERC) under the European Unions Horizon 2020 research and innovation programme (Grant agreement No. 851274). BDS further acknowledges support from an STFC Ernest Rutherford Fellowship.
AK, MCF, and HHo acknowledge support from the Netherlands Organisation for Scientific Research Vici grant 639.043.512.
CH, TT, MA, JHD and BG acknowledge support from  the ERC under grant agreement No.~647112. CH also acknowledges support from the Max Planck Society and the Alexander von Humboldt Foundation in the framework of the Max Planck-Humboldt Research Award endowed by the Federal Ministry of Education and Research and BG from the Royal Society through an Enhancement Award RGF/EA/181006. TT also acknowledges support from the Marie Sk\l{}odowska-Curie grant agreement No.~797794.
MB acknowledges support from the Polish Ministry of Science and Higher Education through grant DIR/WK/2018/12, and the Polish National Science Center through grants no. 2018/30/E/ST9/00698 and 2018/31/G/ST9/03388.
EC acknowledges support from the STFC Ernest Rutherford Fellowship ST/M004856/2 and STFC Consolidated Grant ST/S00033X/1, and from the Horizon 2020 ERC Starting Grant (Grant agreement No 849169).
JD is supported by NSF grant number AST-1814971.
KK acknowledges support from the Alexander von Humboldt Foundation.
SJ is supported by the ERC under grant No.~693024 and the Beecroft Trust.
LM is supported by STFC grant ST/N000919/1.
KM acknowledges support from the National Research Foundation of South Africa.
NS acknowledges support from NSF grant numbers AST-1513618 and AST-1907657.
CS acknowledges support from the Agencia Nacional de Investigaci\'on y Desarrollo (ANID) through FONDECYT Iniciaci\'on grant No.~11191125.
ZX is supported by the Gordon and Betty Moore Foundation.\\
The results in this paper are based on observations made with ESO Telescopes at the La Silla Paranal Observatory under programme IDs 177.A-3016, 177.A-3017, 177.A-3018 and 179.A-2004, and on data products produced by the KiDS consortium. The KiDS production team acknowledges support from: Deutsche Forschungsgemeinschaft, ERC, NOVA and NWO-M grants; Target; the University of Padova, and the University Federico II (Naples).  Data processing for VIKING has been contributed by the VISTA Data Flow System at CASU, Cambridge and WFAU, Edinburgh. \\
For ACT, the work was supported by the U.S. National Science Foundation through awards AST-1440226, AST0965625 and AST- 0408698 for the ACT project, as well as awards PHY-1214379 and PHY-0855887. Funding was also provided by Princeton University, the University of Pennsylvania, and a Canada Foundation for Innovation (CFI) award to UBC. ACT operates in the Parque Astron\'omico Atacama in northern Chile under the auspices of the Comisi\'on Nacional de Investigaci\'on (CONICYT). The development of multichroic detectors and lenses was supported by NASA grants NNX13AE56G and NNX14AB58G. Colleagues at AstroNorte and RadioSky provide logistical support and keep operations in Chile running smoothly. We also thank the Mishrahi Fund and the Wilkinson Fund for their generous support of the project. 
\\
\\
\footnotesize{\textit{Author contributions: All authors contributed to the development and writing of this paper. The authorship list is given in two groups: the lead authors (NCR \& DA \& JHD \& OD \& AK) followed by an alphabetical group of those who made significant contributions to the scientific analysis and/or the ACT or KiDS surveys. }}	

\end{acknowledgements}
\bibliographystyle{aa}
\bibliography{ref}

\input{appendix}
%
%

\end{document}

%% file: appendix.tex
\begin{appendix} 
\section{Intrinsic Alignment Prior}
\label{app}
In this appendix we motivate the choice of prior for the parameter $A_{\rm IA}$, which controls the amplitude of the intrinsic alignment (IA). To do so, we make use of the model presented in \citet{Fortuna2020}, which allows us to predict the IA signal for a KiDS-like survey using parameter constraints inferred from recent IA measurements. Relevant for the discussion here is the treatment of the large-scale alignment signal. The dependence as a function of angular scale $\ell$ is described by the non-linear alignment (NLA) model. The amplitude is computed by distinguishing between red and blue galaxies, where the former dominate the signal in practice. The model assumes that the amplitude does not evolve intrinsically with redshift, but it does depend on the luminosity. As both the mix of red and blue galaxies, as well as the mean luminosity for a flux-limited survey, depend on redshift, this results in a net redshift dependence of the intrinsic alignment signal. 

\begin{figure}[h!]
\centering
\includegraphics[width=\columnwidth]{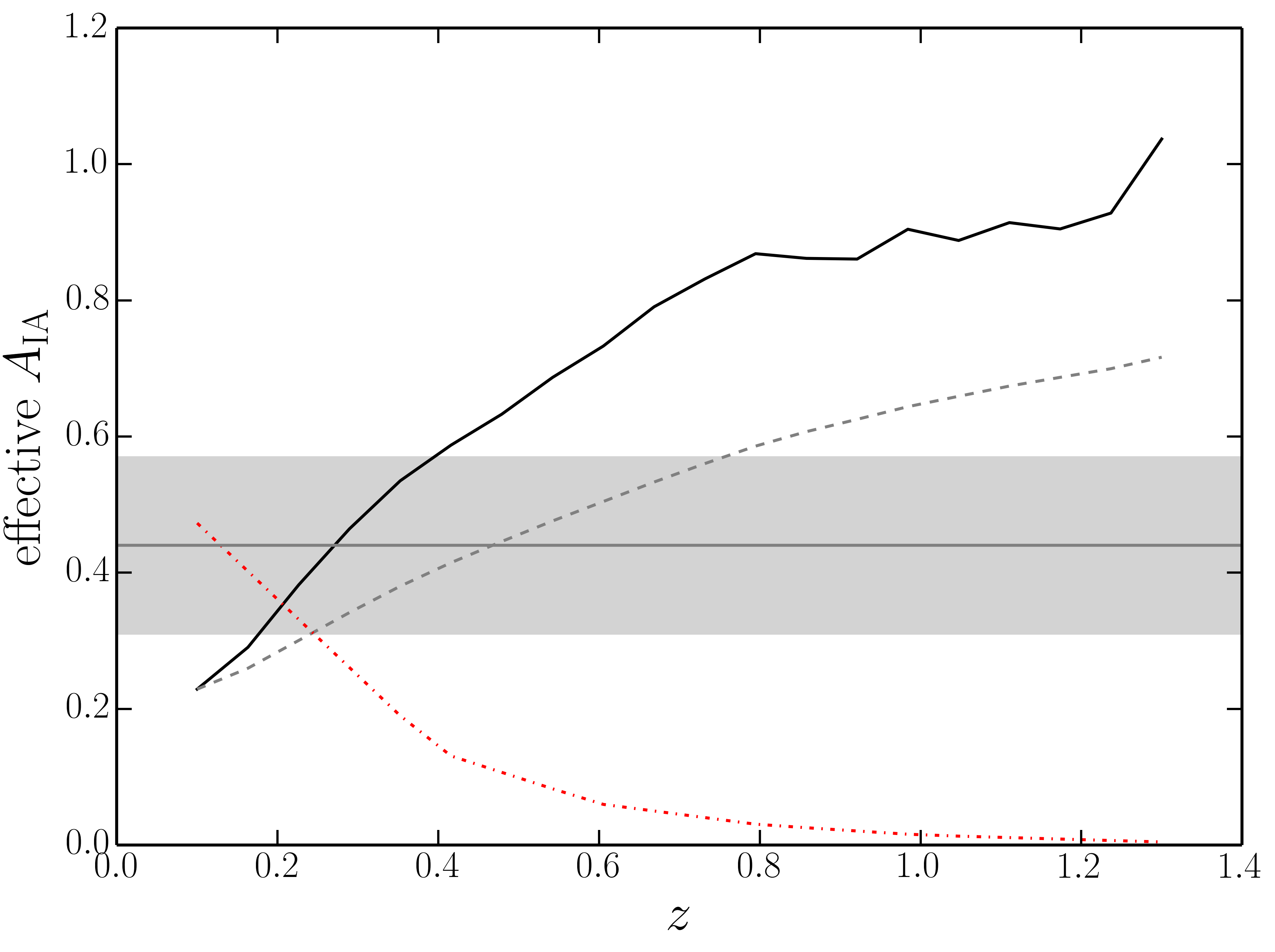}
\caption{The effective amplitude of the large scale intrinsic alignment signal as predicted by model~(ii)
from  \citet{Fortuna2020} (solid black line). The grey dashed lines shows the corresponding cumulative mean amplitude.
Cosmological parameter inference with a constant IA amplitude will result in a low amplitude (grey region; see text for more details), because the relative importance of the IA signal over the cosmic shear signal is higher at low redshift (as indicated by the dashed red line)}
\label{fig:effective_IA_amplitude}
\end{figure}

Observations do not yet constrain the luminosity dependence of the intrinsic alignment signal at the faint end. Consequently, we do not have a good prediction for the typical sources in a cosmic shear survey. To capture the current uncertainty, \citet{Fortuna2020} consider two options: (i) a single power law, which results in a low predicted amplitude; (ii) a broken power law with a shallower slope at low luminosity, leading to a larger net amplitude, because the sources in a lensing survey are generally faint. We show results for the latter in Fig.~\ref{fig:effective_IA_amplitude}.

The solid black shows the predicted net amplitude of the GI signal as a function of redshift for a KiDS-like survey. We do not consider II here, because it does not contribute to a cross-correlation observation. The increase in amplitude is solely caused by the change in galaxy properties. We note that direct observational constraints are largely limited to $z<0.6$; any intrinsic evolution of the alignment signal will change the predictions at high redshift, which are therefore uncertain.

The grey dashed line in Fig.~\ref{fig:effective_IA_amplitude} shows the cumulative mean amplitude. Hence, one would naively expect to infer higher amplitudes as one includes higher redshift sources. However, cosmological parameter inference using a constant IA amplitude yields lower values in practice: the grey band corresponds 
to the best-fit amplitude for a KiDS-like analysis from
\citet{Fortuna2020}. This can be understood from the fact that the relative importance of IA decreases with  redshift. This is illustrated by the red dot-dashed line in Fig. \ref{fig:effective_IA_amplitude}, which shows the mean value of the ratio of $C_\mathrm{IA}^{(ij)}(\ell)/C_\mathrm{GG}^{(ij)}(\ell)$ evaluated at $\ell=1000$ as a function of the redshift of the foreground bin $j$. Hence, even if the actual IA signal increases, it matters less for the best fit parameters, because an error in the IA signal at high redshift has a negligible effect on the inferred $C(\ell)$.

In the case of CMB lensing and galaxy weak lensing cross-correlations we expect the cosmic shear to dominate more, owing to the high redshift of the CMB. In this regard the
grey dashed curve in Fig.~\ref{fig:effective_IA_amplitude}
should be interpreted as an upper limit to the effective IA amplitude to use.

If we naively extrapolate the IA model to $z\sim2$, the cumulative mean amplitude would be $A_\mathrm{IA} \sim 1$. As argued in the previous paragraph, we expect the actual value to use to be lower. Moreover, at such high redshifts we expect the fraction of red galaxies to decline rapidly, as we are approaching the redshift where they are assembled. This provides another argument why we believe the value of $A_\mathrm{IA}$ is lower. 

Given the current uncertainties, it is clear that we cannot use a tight prior on $A_\mathrm{IA}$, but the best fit value from \citet{Fortuna2020} is a reasonable estimate. We therefore 
centre our prior on 0.54, but adopt a wide range so that we span the entire parameter space in the range $A_{\rm IA} \in [0,1]$. With this choice, we cover the scenarios presented in \citet{Fortuna2020}, as well as the current constraints from KiDS cosmic shear analyses \citep{Hildebrandt2020, Wright2020, Asgari2020, Heymans2020}.

\end{appendix}